\documentclass[fleqn,usenatbib,onecolumn]{mnras}

\usepackage{newtxtext,newtxmath}

\usepackage[T1]{fontenc}
\usepackage{ae,aecompl}

\usepackage{graphicx}	
\usepackage{amsmath}	
\usepackage{amssymb}	
\usepackage[dvipsnames]{xcolor}   
\usepackage{comment}

\renewcommand\div{{\mathbf\nabla}{\mathbf\cdot}}

\newcommand\bmb{\boldsymbol{b}}
\newcommand\bmB{\boldsymbol{B}}
\newcommand\bme{\mathbf{e}}
\newcommand\bmE{{\boldsymbol E}}

\newcommand\bmJ{{\boldsymbol{J}}}

\renewcommand\bmu{{\boldsymbol{u}}}
\newcommand\bmv{{\boldsymbol{v}}}

\newcommand\rmd{\mathrm{d}}

\newcommand\rmm{\mathrm{m}}

\newcommand\rmD{\mathrm{D}}
\newcommand\rmM{\mathrm{M}}

\newcommand\f{\frac}
\newcommand\p{\partial}
\newcommand\cst{\mathrm{constant}}

\makeatletter 
\renewcommand*\env@matrix[1][\arraystretch]{%
  \edef\arraystretch{#1}%
  \hskip -\arraycolsep
  \let\@ifnextchar\new@ifnextchar
  \array{*\c@MaxMatrixCols c}}
\makeatother

\title[Magnetic flux transport in protoplanetary discs]{Local semi-analytic models of magnetic flux transport in protoplanetary discs}

\author[Leung \& Ogilvie]{
Philip K. C. Leung$^{1}$\thanks{e-mail: pkcl2@cam.ac.uk}, 
Gordon I. Ogilvie$^{1}$\thanks{e-mail: gio10@cam.ac.uk} 
\\
$^{1}$Department of Applied Mathematics and Theoretical Physics, Centre for Mathematical Sciences, University of Cambridge, \\ 
Wilberforce Road, Cambridge CB3 0WA\\
}

\date{Accepted 2019 June 6. Received 2019 June 3; in original form 2019 April 19}

\pubyear{2019}

\begin{document}
\label{firstpage}
\pagerange{\pageref{firstpage}--\pageref{lastpage}}
\maketitle

\begin{abstract}
The evolution of a large-scale poloidal magnetic field in an accretion disc is an important problem because it determines the launching of winds and the feasibility of the magnetorotational instability to generate turbulence or channel flows. Recent studies, both semi-analytical calculations and numerical simulations, have highlighted the crucial role non-ideal MHD effects (Ohmic resistivity, Hall drift and ambipolar diffusion), relevant in the protoplanetary disc context, might play in magnetic flux evolution in the disc. We investigate the flux transport in discs through the use of two one-dimensional semi-analytic models in the vertical direction, exploring regimes where different physical source terms and effects dominate. The governing equations for both models are derived by performing an asymptotic expansion in the limit of a thin disc, with the different regimes isolated through setting the relative order of the leading terms between variables. Flux transport rates and vertical structure profiles are calculated for a range of diffusivities and disc magnetisations. We found that Ohmic and ambipolar diffusivities drive radially outward flux transport with an outwardly inclined field. A wind outflow drives inward flux transport, which is significantly enhanced in the presence of Hall drift in the positive polarity case, $\eta_H (\boldsymbol{B}_z \cdot \boldsymbol{\Omega}) > 0$, an effect which has only been briefly noted before. Coupled only with outward inclination, the Hall effect reduces the flux transport given by a background Ohmic and/or ambipolar diffusivity, but drives no flux transport when it is the only non-ideal effect present. 
\end{abstract}



\color{black}
\section{Introduction}

Magnetic processes play a significant role in the evolution of accretion discs. Discs are often ionised to some degree \citep{Alexander_etal_2004}, making them susceptible to magnetic effects in the magnetohydrodynamic (MHD) approximation. Magnetic instabilities, such as the magnetorotational instability (MRI), generate turbulence which lead to inward mass accretion by acting as an effective viscosity transporting angular momentum outwards \citep{BalbusHawley1991a}. The presence of a large-scale magnetic field can launch magnetocentrifugal winds (MCWs) which remove angular momentum vertically from the disc surface \citep{BlandfordPayne_1982}, also resulting in mass accretion. Besides these two aforementioned mechanisms which have been subject to extensive studies, magnetic fields have also been theorised to affect disc evolution through other means such as magnetic braking \citep{KrasnopolskyKonigl_2002}, fields linking different radii of the discs \citep{Wardle_2007}, and many more.

A key parameter governing the significance and nature of these processes is the strength and geometry of the large-scale magnetic field threading the disc. The launching of a MCW requires the presence of a large-scale poloidal field above a threshold strength \citep{WardleKonigl_1993}, while the MRI operates most effectively in a `weak field' regime \citep{Fleming_etal_2000}, and is quenched when the flux exceeds a certain value \citep{BalbusHawley_1998}. 

In protoplanetary discs, further complexities arise from the presence of non-ideal MHD effects (Ohmic resistivity, ambipolar diffusion and the Hall drift) from their weak ionisation level. These provide additional parameters affecting the launch criterion of the MCW \citep{WardleKonigl_1993}, and whether or not the MRI is stabilised \citep{Wardle1999,BalbusTerquem_2001,KunzBalbus_2004,SalmeronWardle_2005}. There is growing evidence that non-ideal MHD effects, when coupled with a large-scale magnetic field, can induce the formation of self-organised structures such as axisymmetric rings \citep{Bethuneetal2016,Bethuneetal2017,Suriano_etal_2018}. Such features have been observed in many protoplanetary systems \citep{ALMA_2015} and are important for the study of planet formation.

So far, two possible origins for this large scale magnetic field have been proposed: it is either created locally by a dynamo process \citep{Brandenburg_etal_1995,HGB_1996}, or it arises from the concentration of the flux that is intrinsic to the accreting gas \citep{Spruit_etal_1995,LubowSpruit_1995}. It is yet unclear whether the former can generate a significant magnetic field coherent over a scale comparable to the radius \citep{Spruit_2010}. 

The latter case, where magnetic flux of the accreting gas is concentrated as it accumulates in the inner regions of the disc, has been subject to increasing study. Semi-analytic work such as carried out by \citet{Lovelace_etal_2009} and \citet{GuiletOgilvie2012,GuiletOgilvie2013} studied the competing effects of inward accretion due to advection with an accreting flow, and outward diffusion due to a turbulent resistivity, in what is known as the `advection-diffusion' paradigm. The disc is assumed to be turbulent due to processes like the MRI. This, as mentioned before, leads to an effective viscosity, $\nu$, which drives accretion of the material towards the star, also advecting magnetic flux radially inwards. However, the same turbulence also mixes up fluid parcels and their magnetic flux, driving reconnection processes on small scales. This leads to an effective resistivity, $\eta$, which allows the magnetic field to diffuse, and in the presence of an outwardly bending global poloidal field, results in an outward transport of flux. The turbulent viscosity and diffusivity are linked via the effective magnetic Prandtl number $\mathcal{P}\equiv \nu/\eta$. For MHD turbulence, $\mathcal{P}$ is expected to be of order unity \citep{FromangStone_2009,GuanGammie2009,LesurLongaretti_2009}.

The first studies by \citet{Lubow_etal_1994a} and \citet{Heyvaerts_etal_1996} concluded that outward diffusion would significantly outweigh any accretion for thin discs with significant field bending. These studies, however, neglected the vertical structure of the disc, which, as subsequently pointed out \citep{OgilvieLivio2001,RothsteinLovelace_2008,GuiletOgilvie2012}, has a significant impact on the averaging procedure used in the calculations. These studies examined the impact of disc structure using quasi-steady state radially local models, and found that for weak magnetic fields, the contribution of the accretion flow to the flux transport velocity is much larger than the mass transport velocity. This is due to larger radial velocities away from the mid-plane, which barely affect mass flow but do affect flux transport. Further work by \citet{GuiletOgilvie2014, Okuzumi_etal_2014} and \citet{TakeuchiOkuzumi_2014} applied the results from these local models to the global evolution of large-scale magnetic fields, and verified that an equilibrium is reached and that the maximum attainable field strength is in agreement with the steady-state accretion rate observed in actual systems. Hence a possible solution to the too-efficient-diffusion problem of an inclined magnetic field is provided.

This flux transport paradigm, however, is more relevant to well ionised discs such as around black holes and compact stars, where turbulence can readily occur. Recent studies and global simulations have shown that PPDs are likely to be laminar in most parts, and therefore should not be modeled as turbulent \citep{Bethuneetal2017,BaiStone2017,Suriano_etal_2017,Suriano_etal_2018}. Current advection-diffusion models also do not account for the non-ideal effects of Hall drift and ambipolar diffusion, which are likely to be significant throughout much of the PPD. Recently numerical simulations have begun examining the flux transport problem globally in the non-ideal MHD protoplanetary disc context \citep{BaiStone2017} and in the ideal MHD weak field regime \citep{ZhuStone2018}. However, they still suffer from current computational limitations of evolving a disc for long enough to determine its long term global flux evolution (G. Lesur, priv. comm.). A new model addressing these concerns and including all three non-ideal MHD contributions is required for modelling the flux transport problem for the PPD context, which is what we seek to address in this work.

In this paper, we investigate the flux transport problem in the PPD context through 1D radially local semi-analytic models similar to that of \citet{GuiletOgilvie2012}, but where non-ideal effects also have a significant role in the flux evolution \citep{Fromang2013}. However, as many of the same principles can be applied to accreting discs in general, it is hoped that the framework could be extended to the study of stellar mass black hole accretion and AGN discs as well.

The structure of this paper is as follows. In Section 2, we construct the multiscale aysmptotic approach which forms the basis for the two semi-analytic models used in this work. Section 3 discusses the shearing box equations vertical structure model, where we estimate flux transport rates due to an inclined $B$ field or a magnetic torque at the disc surface. Section 4 extends the \citet{GuiletOgilvie2012} model (hereafter GO model) to include all three non-ideal effects. It also examines the flux transport driven by large scale radial gradients, in addition to that by a surface inclined field or magnetic torque. The implications and limitations of our results are discussed in Section 5, along with directions for future study. Finally, we summarise and conclude in Section 6.

\section{A multiscale asymptotic approach}
\label{sec:Multscale}

\citet{GuiletOgilvie2012} were distinctive in using a multiscale asymptotic approach in deriving their 1D radially local vertical structure equations, allowing radial gradients of various quantities to be placed on the same footing as inclination and outflow in driving flux transport. This approach is often employed in the study of warped discs where processes vary at different spatial and time scales \citep{Ogilvie_1999}.
We follow the same multiscale asymptotic approach, but here derive a more general formalism. All our subsequent work is based on the equations derived using this formalism, and here we use the equations derived to also draw some general observations on the interplay between various source terms and the angular momentum transport and magnetic flux evolution. 

\subsection{Governing equations}

The continuity equation is given by
\begin{equation}
\p_t \rho + \nabla\cdot ( \rho \bmu ) = 0,
\end{equation}and the conservation of momentum by
\begin{equation}
    \rho \left( \p_t \bmu + \bmu \cdot \nabla \bmu \right) = - \rho \nabla \Phi - \nabla p + \f{1}{\mu_0} (\nabla\times\bmB) \times \bmB + \nabla\cdot\boldsymbol{\mathrm{T}},
\end{equation}
where $\rho$ is the density, $\bmu$ is the velocity,  $\Phi$ is the gravitational potential, $p$ is the pressure, $\bmB$ is the magnetic field, and $\mathbf{T}$ is the viscous stress. The full viscous stress is given by
\begin{equation}
    \boldsymbol{\mathrm{T}} = \rho\nu 
    \left[ \nabla \bmu + (\nabla \bmu)^{\mathrm{T}}
    - \f{2}{3}(\nabla\cdot\bmu)\boldsymbol{\mathrm{I}} \right],
\end{equation}
where $\nu$ is the kinematic viscosity and $\boldsymbol{\mathrm{I}}$ is the unit tensor of second rank. We neglect self-gravity and do not consider a thermal energy equation. For thermodynamic closure, we assume an isothermal relation such that 
\begin{equation}
p = c_s^2 \rho.
\end{equation}

The full non-ideal MHD induction equation is given by \citep{Balbus_2011}:
\begin{equation}\label{eq:fullinduct}
    \begin{aligned}
        \partial_t \boldsymbol{B}
         = & \nabla \times \bigg( \bmu\times\bmB
        - \eta_O(\nabla\times\bmB)  - \eta_H (\nabla\times\bmB) \times \bmb \\
        & \qquad\qquad
        +\eta_A \left[ (\nabla\times\bmB)\times\bmb\right] \times \bmb \bigg),
    \end{aligned}
\end{equation}
where $\bmb = \bmB/\lvert\bmB\rvert$, and $\eta_O, \eta_H, \eta_A$ are the Ohmic, Hall and ambipolar diffusivities respectively. Ohmic and ambipolar diffusivities are always positive. The Hall coefficient, on the other hand, can have either sign, depending on whether the more massive charge carrier in the disc is positively ($\eta_H>0$) or negatively ($\eta_H<0$) charged. Ideal, Ohmic and/or ambipolar diffusive only discs are insensitive to the polarity of alignment of the large scale field with the rotation of the disc. The Hall effect, on the other hand, does depend on the polarity of the magnetic field. Reversing the sign of $(\boldsymbol{B}_z\cdot \boldsymbol{\Omega})$ has the equivalent effect as reversing the sign of $\eta_H$ on the equations. In this paper, we define the positive (negative) polarity case as when $\eta_H (\boldsymbol{B}_z \cdot \boldsymbol{\Omega}) > (<)~ 0$. 

We consider the problem in cylindrical polar coordinates $(r,\phi,z)$, and assume an axisymmetric potential $\Phi(r,z)$ that is symmetric about the midplane $z=0$. The orbital angular velocity $\Omega(r)$ is defined via
\begin{equation}
  -r\Omega^2=-\p_r\Phi_\rmm,
\end{equation}
where
\begin{equation}
  \Phi_\rmm(r)= - \f{GM}{r}
\end{equation}
is the midplane potential.  We also define the residual velocity as
\begin{equation}
  \bmv=\bmu-r\Omega\,\bme_\phi,
\end{equation}
which is the departure from Keplerian motion.

\subsubsection{Asymptotic expansion}

We perform an asymptotic expansion for the fluid variables, using an ordering scheme that is well understood for thin discs. The characteristic angular semi-thickness of the disc is given by the disc aspect ratio $H/r = \mathcal{O}(\epsilon)$, where $0<\epsilon\ll 1$ is a small dimensionless parameter. The sound speed to orbital velocity ratio in an isothermal disc is then of $\mathcal{O}(\epsilon)$.  

For a magnetised disc, the ordering scheme of the magnetic field depends on the strength and orientation of the field \citep{Ogilvie_1997}. We consider here a situation where the magnetic energy density is comparable to the thermal energy density, and set both the Alfv\'en speed and the adiabatic sound speed at $\mathcal{O}(\epsilon)$ compared with the orbital velocity. For specific field configurations, such as vertical field dominance in the Guilet \& Ogilvie model described in Section \ref{sec:GuiletOgilvie}, we might set some of the asymptotic terms to zero for that particular situation. We neglect the effects of a viscous stress, setting the viscosity to zero, giving us a laminar disc, as recent PPD simulations have shown them to be.

We assume axisymmetry and so neglect all $\p_\phi$ terms. The internal structure of the thin disc and its slow evolution in time are resolved through rescaled spatial and time coordinates
\begin{equation*}
  \xi=\f{r}{\epsilon},\qquad
  \zeta=\f{z}{\epsilon},\qquad
  t_1 = \epsilon t,\qquad
  t_2 =\epsilon^2t,
\end{equation*}
where $\eta$ is a scaled angular variable, $t_1$ is the flux evolution timescale and $\tau$ is the accretion timescale. To incorporate the multiscale nature of our approach, we separate the radial lengthscale into small variations of order $H$ within a small radial region of the disc ($\xi\sim\mathcal{O}(1)$) and global radial variations of order $r$ ($r\sim\mathcal{O}(1)$). Timescales are also separated into a fast dynamical timescale of order $\Omega^{-1}$ ($t\sim\mathcal{O}(1)$), an intermediate magnetic flux transport timescale ($t_1\sim\mathcal{O}(1)$), and a slow accretion timescale ($t_2\sim\mathcal{O}(1)$). The coordinate transformations are hence given by
\begin{eqnarray*}
  &&\p_r\mapsto\epsilon^{-1}\p_\xi+\p_r,\qquad
  \p_z\mapsto\epsilon^{-1}\p_\zeta,\nonumber\\
  && \p_t\mapsto \p_t+\epsilon\p_{t_1}+\epsilon^2\p_{t_2}.
\end{eqnarray*}

We then use the following expansion of fluid variables (each of the terms in the expansion is a function of $r,\xi,\zeta,t,t_1,t_2$):
\begin{eqnarray}
  &&\rho=\rho_0+\epsilon\rho_1+\cdots, \\
  &&p=\epsilon^2\left[p_0+\epsilon p_1+\cdots\right], \\
  &&\bmv=\epsilon\left[\bmv_0+\epsilon\bmv_1+\cdots\right], \\
  &&\boldsymbol{B} = \epsilon\left[\boldsymbol{B}_0 + \epsilon \boldsymbol{B}_1 + \cdots\right],\\
  &&\eta_O=\epsilon^2[\eta_{O0}+\epsilon\eta_{O1}+\cdots], \\
    &&\eta_H=\epsilon^2[\eta_{H0}+\epsilon\eta_{H1}+\cdots], \\
    &&\eta_A=\epsilon^2[\eta_{A0}+\epsilon\eta_{A1}+\cdots].
\end{eqnarray}
We can see that $p/\rho \sim \mathcal{O}(\epsilon^2)$, giving us $c_s\sim\mathcal{O}(\epsilon)$ as we would expect. We consider departures from Keplerian motion comparable to the sound speed, and a similar order for the Alfv\'en speed, hence set both $\bmv$ and $\bmB$ to be of $\mathcal{O}(\epsilon)$. Variations across the disc height due to diffusive effects are assumed to be at the dynamical timescale, hence $\eta_{O,H,A}\sim\mathcal{O}(\epsilon^2 r^2/t)\sim\mathcal{O}(\epsilon^2)$. Finally, within the thin disc, the gravitational potential has the Taylor expansion
\begin{equation}\label{eq:multiscale.potential}
  \Phi=\Phi_\rmm(r)+\epsilon^2\Psi(r)\f{1}{2}\zeta^2+O(\epsilon^4),
\end{equation}
where
\begin{equation}
  \Psi=\Omega^2
\end{equation}
for the point-mass potential we have chosen.

At leading order ($\mathcal{O}(1)$ for mass conservation and $\mathcal{O}(\epsilon)$ for the momentum equation), we recover the standard shearing box equations. The next order equations can be thought of as linear equations with source terms given by the solution of the leading order equations. The expansion of these equations at the first two leading orders can be found in Appendix A (see online supplementary materials).

\subsubsection{Angular momentum and magnetic flux transport}
\label{sec:multiscale.analysis.fluxtransport}

Of particular interest to us are the leading terms for angular momentum and magnetic flux transport rates. Assuming that disc variables are periodic in $\xi$, we simplify the equations by averaging over the horizontal directions and over the fast dynamical timescale $t$. The leading order angular momentum transport equation is then recovered by combining the $\mathcal{O}(\epsilon^2)$ $\phi$-component of the momentum equation with the $\mathcal{O}(\epsilon)$ mass equation, and integrating over the vertical extent with no outflow boundary condition to give
\begin{equation}
\begin{aligned}
  \p_\tau \Sigma_0
  = &\f{1}{r}\p_r\bigg\{
  \left[\p_r(r^2\Omega)\right]^{-1} \\
  & \qquad\qquad
  \left[\p_r\left(r^2 \left\{ \langle \rho_0v_{\phi0}v_{r0} \rangle 
  - \f{1}{\mu_0} \langle B_{r0} B_{\phi0}  \rangle \right\} \right)
   \right]\bigg\},
\end{aligned}
\end{equation}
where $\langle\rangle$ denote the spatial and temporal averagings and the vertical integration mentioned, and $\Sigma_0$ is the first order disc surface density.
We recover the well known `diffusion equation' for an accretion disc, that mass transport is facilitated through
`viscous' stresses arising from a Reynolds component associated with internal motions ($\langle \rho_0v_{\phi0}v_{r0} \rangle$) and a Maxwell component associated with the horizontal magnetic field components ($-\langle B_{r0} B_{\phi0}  \rangle$).

\begin{figure}
    \begin{center}
	\includegraphics[width=0.5\columnwidth]{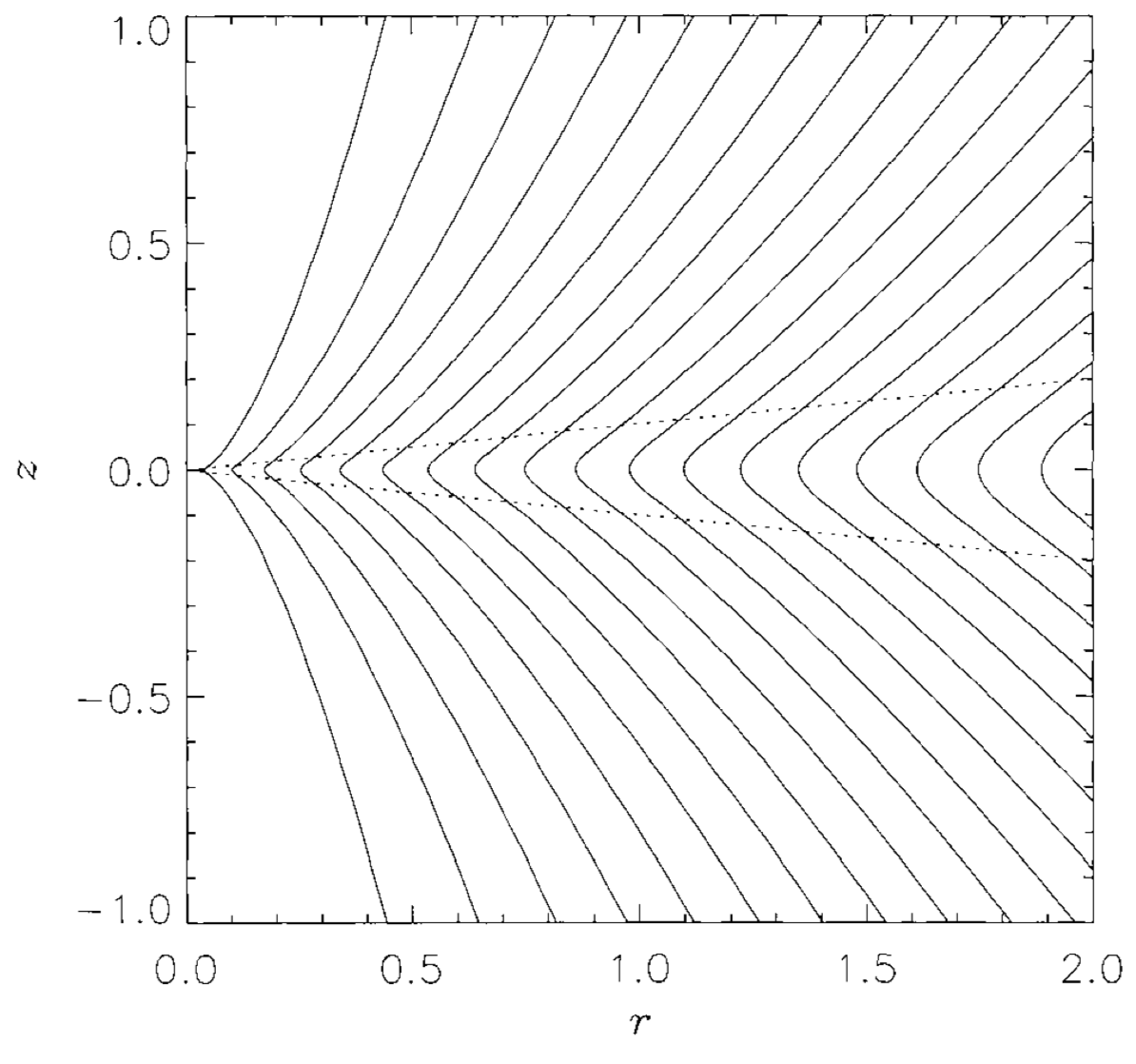}
    \end{center}
    \caption{General form of the open disc magnetic field geometry assumed in standard accretion disc models. Magnetic field lines are shown in the meridional plane, while the dotted lines indicate the surfaces of the disc. Figure taken from \citet{Ogilvie_1997}.}
    \label{fig:standardBconfig}
\end{figure}

The vertical flux evolution is calculated from the $z$ component of the induction equation and using the condition $\nabla\cdot \bmB = 0$:
\begin{equation}\label{Bzfluxevolution_ep2NonIdeal}
\setlength{\jot}{10pt}
\begin{aligned}
    &\p_{t_1} \langle B_{z0}\rangle+\f{1}{r}\p_r\bigg(r\bigg[\langle v_{r0}B_{z0}\rangle
    -\langle v_{z0} B_{r0}\rangle
    +\left\langle\eta_{O0}\p_\zeta B_{r0}\right\rangle\\
    &\qquad
    +\bigg\langle\eta_{H0} \f{1}{B_0}
    \bigg\{ B_{r0}\p_\xi B_{\phi0}
    + B_{z0}\p_\zeta B_{\phi0}\bigg\}\bigg\rangle\\
    &\qquad
    +\bigg\langle\eta_{A0} \f{1}{B_0^2}
    \bigg\{ B_{r0} B_{\phi0}\p_\zeta B_{\phi0} 
    + \left(B_{r0}^2 + B_{z0}^2\right)\left(\p_\zeta B_{r0} - \p_\xi B_{z0} \right)\\
   &\qquad\qquad
   + B_{\phi0} B_{z0} \left(
    \p_\xi B_{\phi0}
  \right)
    \bigg\}\bigg\rangle \
    \bigg]\ \bigg)=0.
\end{aligned}
\end{equation}
The terms within the square brackets give us the vertical magnetic flux radial transport rate. The first term describes advection with the radial flow, while the second term shows us the effect from a vertical outflow from the disc. Standard accretion disc models assume the large scale magnetic field to have an open field geometry with the poloidal component bending outward from the star \citep{Ogilvie_1997} (see Figure \ref{fig:standardBconfig}). In this configuration, a vertical outflow would therefore drive radially inward accretion of the flux ($-\langle v_{z0} B_{r0}\rangle<0$). For the Ohmic term, we expect $\langle \p_\zeta B_{r0} \rangle > 0$ in the standard disc magnetic configuration. This means an overall effect of diffusing the magnetic field radially outwards, in agreement with other studies of the effect of Ohmic resistivity in the presence of an inclined field \citep{GuiletOgilvie2012}. 

The Hall and ambipolar terms are more complicated. We further simplify the problem by assuming that only large scale radial gradients and vertical variations are present (in other words we set $\p_\xi=0$). Then the Hall term, 
\begin{equation}\label{eq:multiscale.hall}
    \left\langle \eta_{H0} \left( \f{ B_{z0}}{B_0} \right) \p_\zeta B_{\phi0} \right\rangle ,
\end{equation}
depends on the vertical gradient of the azimuthal field. Its direction is also dependent on the sign of the Hall coefficient and the vertical magnetic field, reversing if one of them changes. Its effect is strongest when there is a strong vertical field threading the disc. From simulations \citep{BaiStone2017,Bethuneetal2017}, $\p_\zeta B_{\phi0} < 0$ near the midplane and $>0$ at the disc surface, hence we would expect the Hall flux transport to switch direction as we go up from the disc midplane. For $\eta_H (\boldsymbol{B}_z \cdot \boldsymbol{\Omega}) > (<)~ 0$, we expect this transport to be radially inward (outward) near the midplane. This is indeed what is found in recent Hall global simulations \citep{BaiStone2017}. 

The ambipolar term on the other hand has both an `Ohm-like' and a `Hall-like' component. The `Ohm-like' component,
\begin{equation}\label{eq:multiscale.amb.ohmlike}
    \left\langle \eta_{A0} \left( \f{B_{r0}^2 + B_{z0}^2}{B_0^2} \right) \p_\zeta B_{r0} \right\rangle ,
\end{equation}
is outwardly directed assuming the standard configuration, and is strongest and equivalent in form to Ohmic diffusion when $\bmB$ is purely poloidal. The `Hall-like' component,
\begin{equation}\label{eq:multiscale.amb.halllike}
    \left\langle \eta_{A0} \left( \f{ B_{r0} B_{\phi0}}{B_0^2} \right) \p_\zeta B_{\phi0} \right\rangle ,
\end{equation}
depends on the correlation between the horizontal components of the magnetic field. Normally, we expect the sign of $B_\phi$ to be opposite to that of $B_r$ because of Keplerian shear, and $B_{r0} B_{\phi0} < 0$. Hence the `Hall-like' ambipolar term would act in the opposite (same) direction as the Hall term for $\eta_H (\boldsymbol{B}_z \cdot \boldsymbol{\Omega}) > (<)~ 0$.

It is worth noting that Section 2 (and Figure 1) of \citet{BaiStone2017} also provides an analytic explanation on why the Hall term drives magnetic flux transport when a surface toroidal field accompanied by a wind is present, with Equation 9 of \citet{BaiStone2017} essentially the same as our Equation \eqref{eq:multiscale.hall}. Although our analysis is more rigorous, the essence of the Hall-mediated magnetic flux transport is the same.

\section{A shearing box (radially local) 1D vertical structure model}
\label{sec:RLVSM}

In our first investigation, we developed a semi-analytic radially local 1D vertical model that includes all non-ideal MHD effects for the flux transport problem. Our first model is essentially a local shearing box, where we solve the leading order equations, and assume variations only in the vertical direction. This assumption is motivated by global non-ideal MHD simulations such as those of \citet{BaiStone2017}, where quasi-steady states were achieved and the 1D vertical disc profiles were measured. The goal of our approach is to gain insight into the interpretation of their results.

We begin by examining the case with no MCW present by restricting the range of field inclinations explored to below the wind launching condition, but then mimic the presence of a vertical outflow by setting a non-zero azimuthal field, hence magnetic torque, at the boundary. In our model, flux transport is driven by the interplay between the bending of the poloidal field and the diffusivities present.

\subsection{Leading order equations}
\label{sec:RLVSM.model}

We use the thin disc approximation ($J_z=0$, see Equation \eqref{RLVSM:nonideal}), and consider a Keplerian rotating frame.
We want to find the quasi-steady state equilibrium profiles, so the disc is set to be steady on the dynamical timescale (i.e. $\p_t=0$). As before, we assume an isothermal and inviscid flow. For now, there is also no vertical outflow, so we set $v_z=0$.

Without loss of generality, we use units such that $B_z, \Omega, c_s, \mu_0 = 1$, with $B_z$ set to constant in this model. The vertical coordinate $z$ is then in units of the isothermal scaleheight, $H=c_s/\Omega$, while  diffusivities are given in units of $H^2 \Omega$. The relation between the Alfv\'en speed and the sound speed (a measure of the field strength) is then defined by our choice of the midplane ($z=0$) density, $\rho(0)$:
\begin{equation}
\begin{aligned}
    c_s=1\implies v_a/c_s=v_a,
\end{aligned}
\end{equation}
\begin{equation}
\begin{aligned}
    v_a=\f{B}{\sqrt{\mu_0\rho}}=\sqrt{\f{B_x^2+1}{\rho}},\qquad
    v_a(0)=\sqrt{\f{1}{\rho(0)}}.
\end{aligned}
\end{equation}
$\rho(0)\gg 1$ therefore corresponds to a weak magnetic field and $\rho(0)\ll 1$ to a strong one. When $\rho(0)=1$, $c_s=v_a$ at the midplane. For an isothermal disc profile, $\rho(z)$ decreases as $z$ increases, so we expect magnetic effects to become stronger compared to hydrodynamic effects as we move away from the midplane. 

The local equations are identical to the shearing box equations, and correspond to the leading order equations of the multiscale asymptotic approach. We consider only vertical variations, setting $\p_\xi=0$. To incorporate non-ideal effects, we use the modified electric field in the rest frame of the multi-component fluid $(\bmE + \bmv\times\bmB)$, with $\bmE$ being the electric field in the shearing-box frame:
\begin{equation}\label{RLVSM:nonideal}
    \bmE + \bmv\times\bmB = \eta_O\bmJ
    + \eta_H\bmJ\times\bmb - \eta_A(\bmJ\times\bmb)\times\bmb,
\end{equation}
where $\bmJ=\nabla\times\bmB$ is the current density, $\bmv$ throughout this section is the fluid velocity (rather than the Keplerian-shear subtracted velocity in Section \ref{sec:Multscale}), and the other variables are as before. This is another way of writing the modified induction equation, since $\p_t \bmB = - \nabla \times \bmE$. In our scheme here, the positive (negative) polarity configuration for the Hall effect is achieved when $\eta_H> (<)~ 0$. 

We then have the following governing equations:
\begin{equation}\label{RLVSM1.1}
    -2\rho\Omega v_{y} = J_{y} B_{z},
\end{equation}
\begin{equation}\label{RLVSM1.2}
    \f{1}{2}\rho\Omega v_{x} = - J_{x} B_{z},
\end{equation}
\begin{equation}\label{RLVSM1.3}
    0 = - \rho \Omega^2 z - \f{\rmd p}{\rmd z} + J_x B_y - J_y B_x.
\end{equation}
\begin{equation}\label{RLVSM1.4}
    0 = \f{\rmd E_y}{\rmd z},
\end{equation}
\begin{equation}\label{RLVSM1.5}
    0 = -\f{3}{2}\Omega B_x - \f{\rmd E_x}{\rmd z},
\end{equation}
with $x,y,z$ denoting the radial, azimuthal and vertical directions respectively, and we have removed the ordering subscripts. Equations \eqref{RLVSM1.1} to \eqref{RLVSM1.3} are the three components of the equation of motion, while the horizontal components of the induction equation are given by Equations \eqref{RLVSM1.4} and \eqref{RLVSM1.5}.
These terms on the RHS of Equations \eqref{RLVSM1.1} and \eqref{RLVSM1.2} indicate that the departure from Keplerian motion are driven by the Maxwell terms. We can also see that the density structure is determined by the effects of gravity, pressure balance, and magnetic compression, which are represented by the first, second and final pair of terms respectively of Equation \eqref{RLVSM1.3}.

\subsection{Flux transport}

The flux transport rate is given by the azimuthal electric field $E_y$, which is constrained to be a constant with varying disc height in our model. We can see that it measures the flux transport rate by examining the vertical component of the induction equation in cylindrical geometry:
\begin{equation}\label{eq:inductvert}
    \p_t B_z = - \f{1}{r}\p_r (r E_\phi).
\end{equation}
In strict steady state, the left hand side of Equation \eqref{eq:inductvert} is zero. But since $B_z$, and equivalently the poloidal magnetic flux function $\Psi(r,z)$ (see Sections 2 \& 3.3 in \citet{GuiletOgilvie2012} for a justification using that approach), can evolve on the long accretion time-scale $\tau_a\equiv r/\lvert v_r \rvert $ as the poloidal field drifts radially through the disc, we can define a radial flux transport rate to be given by
\begin{equation}\label{eq:RLVSM.vPsi.def}
    v_{\Psi} = E_y/B_z
\end{equation}
\citep{OgilvieLivio2001,Konigl_etal_2010,GuiletOgilvie2012}. This is found in our model as an eigenvalue of the solution. $E_y > (<)~ 0$ would hence imply a radially outward (inward) transport. 

\subsection{Boundary conditions and numerical method}
\label{sec:shooting}

We use the standard assumption that the disc is antisymmetric about the midplane in $B_x$ and $B_y$, and symmetric in $v_x$ and $v_y$. The value of $B_x(z\to\infty)$ sets the inclination of the poloidal field, which is one of the parameters we vary. In a real disc, the inclination would be determined by the global magnetic field geometry. For the case where there is no wind launch, we expect no external magnetic torque to be acting on the disc, and set $B_y=0$ at $z\to\infty$. In the case where we mimic the presence of a wind, we set $B_y$ to be non-zero at $z\to\infty$, since there is now an external magnetic torque acting on the disc removing angular momentum vertically. For an actual outflow, we would expect $B_y(\infty)<0$. The value of $\rho(z=0)$ is another parameter we vary and sets the field strength. In our units, the midplane density $\rho_0$ is equivalent to $\beta_0/2$, where $\beta = p_\text{thermal}/p_\text{magnetic}$ is the plasma beta. A large value of $\rho_0$ would therefore indicate a weak midplane magnetic field.

We recast Equations \eqref{RLVSM1.1}$-$\eqref{RLVSM1.5} into a fourth order ordinary differential system with variables $\rho$, $B_x$, $B_y$ and $E_x$ (see Appendix B in the online supplementary materials for details), where $E_y$ becomes an eigenvalue of the problem, and solve them using the shooting method. We integrate upwards from the midplane with values determined by the boundary conditions and guessed midplane values for $E_x$ and $E_y$. The ODE system is then integrated using fourth order Runge-Kutta up to a sufficient height to mimic the solution as $z\to\infty$ (we set $z_\text{end} = 10 H$ in our case, where $H$ is the isothermal scale height. Extending $z_\text{end}$ beyond this value is found to yield no significant variation in our results). There, boundary conditions at infinity are applied. This is followed by Newton-Raphson iterations to adjust the guessed solution until the variables converge to the desired upper boundary conditions.

\subsubsection{Diffusivity profiles}

We used two different diffusivity profiles in our calculations. The first profile, `Uniform', assumes uniform $\eta_O$, $\eta_H$ and $\eta_A$ across all scale heights, and is useful for determining the general behaviour of the disc due to each diffusivity. The second scheme, `CstIon', tries to capture PPD vertical diffusivity profiles more accurately by assuming only constant ionisation across the vertical extent, leading to  $\eta_O = \cst$, $\eta_H\propto B/\rho$ and $\eta_A \propto B^2/\rho^2$ \citep{Wardle_2007,Balbus_2011}. We also impose a rapid transition to the ideal MHD regime for $z>2$ to model the effect of photoionisation of the disc surface from the star. In all the plots that follow, the labelled values of  $\eta$ are true for the whole disc in the `Uniform' case, whereas for the `CstIon' case, they are the values of the midplane diffusivities. 

\subsection{Marginal stability analysis}
\label{sec:RLVSM.MargStab}

According to \citet{Ogilvie_1998}, steady state solutions where field lines bend several times as they pass through the disc indicate their instability to the MRI in the ideal MHD regime. Although it yet remains to be rigorously proven, we follow \citet{OgilvieLivio2001} and \citet{GuiletOgilvie2012} in assuming that this to be true when non-ideal effects are included as well. The multiple bending corresponds to the ``channel mode'' of the MRI for a vertical field. The lowest order ``$n=1$'' mode, which determines the overall marginal stability, has opposite symmetry in the disc to the fluid variables. There, $\Delta B_x$ and $\Delta B_y$, where $\Delta$ denotes a perturbation, would be symmetric about the midplane, while $\Delta u_x$ and $\Delta u_y$ would be antisymmetric. Hence we can compute the level of magnetisation (in our case given by $\rho_0=\beta_0/2$) that leads to a marginally stable state to the MRI by solving for a set of linearised equations of small perturbations on top of the nonlinear equations, with the appropriate symmetry conditions.

This was done for a range of diffusivity coefficients using the shooting method described in Section \ref{sec:shooting}. Midplane values of $\Delta B_y$ and $\rho$ were guessed along with $E_x$ and $E_y$ as before ($\Delta E_x$ is calculated as an output, while $\Delta E_y$ is zero due to the symmetry of the perturbation). The midplane value of $\Delta B_x$ determines the amplitude of the marginal mode, and was arbitrarily set to $10$. Boundary conditions for the unperturbed variables were the same as before. For the perturbed variables, midplane values are derived from the symmetry conditions ($\Delta u_x=\Delta u_y=0$), while the upper boundaries are set to enforce the fixed inclination ($\Delta B_x=0$) and magnetic torque ($\Delta B_y=0$) conditions. Threshold field strengths are then computed, which mark the field strengths below which the disc would be unstable to the MRI, and where we would expect multiple bending in the vertical structure solutions.

\subsection{Disc vertical structure profiles}

\begin{figure}
    \begin{center}
	\includegraphics[width=0.3\columnwidth]{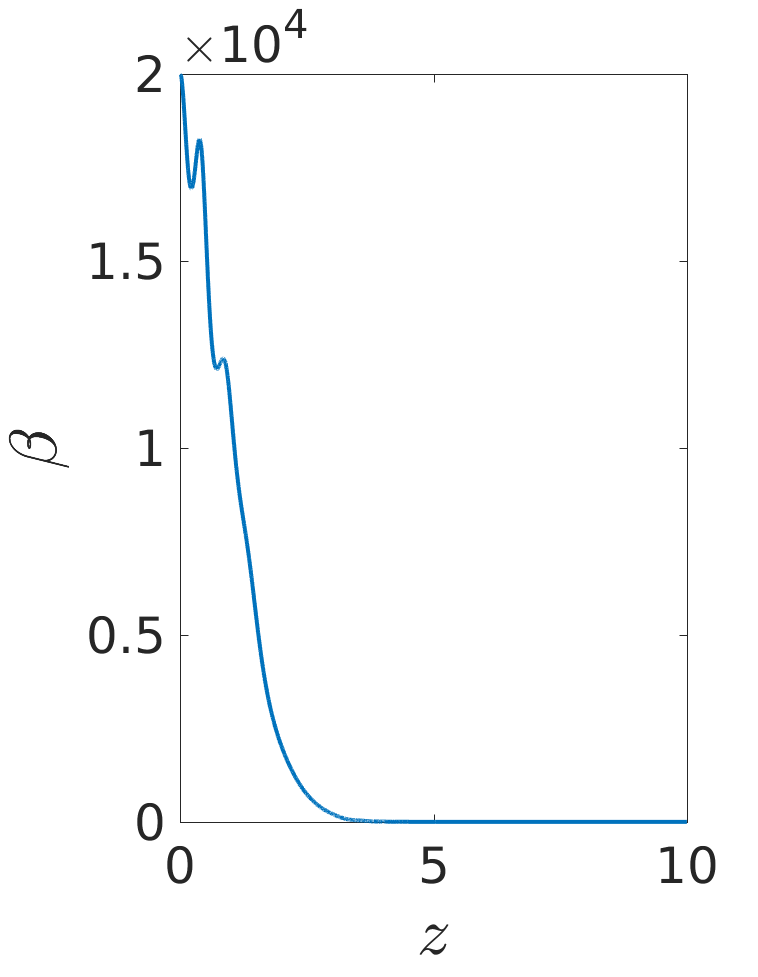}
	\includegraphics[width=0.3\columnwidth]{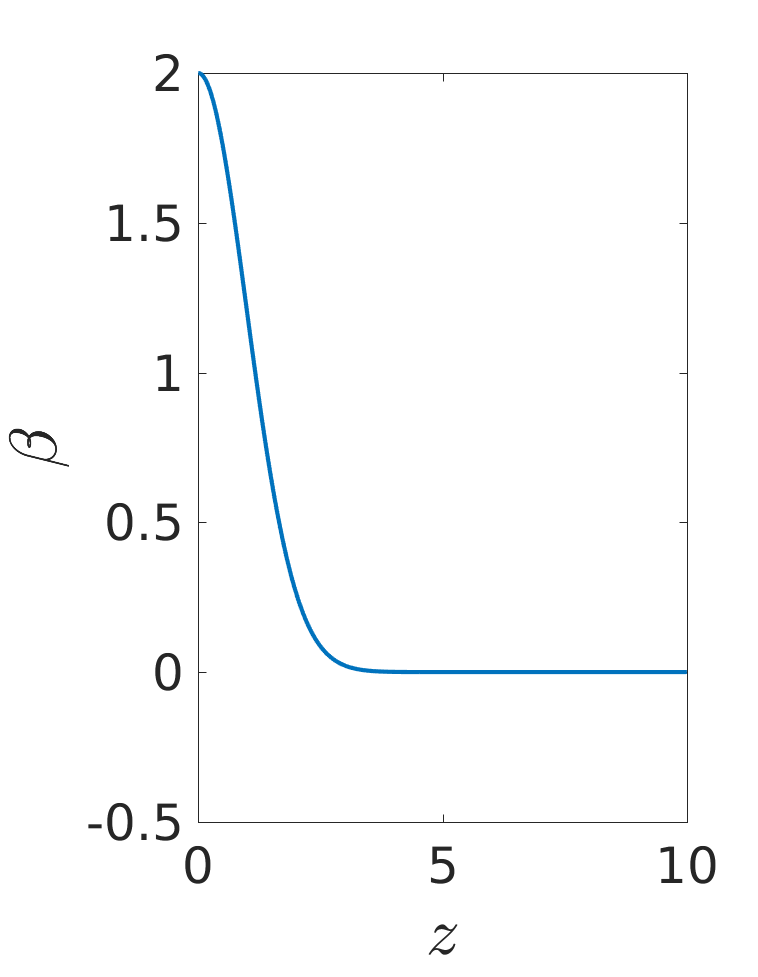}\\
	\includegraphics[width=0.3\columnwidth]{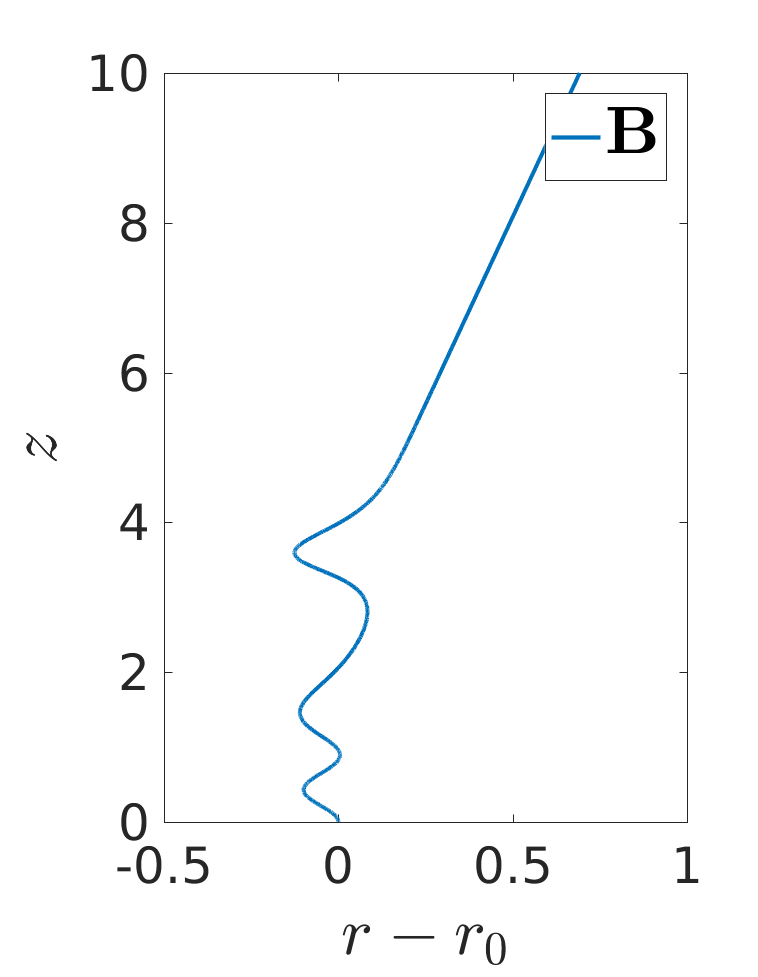}
	\includegraphics[width=0.3\columnwidth]{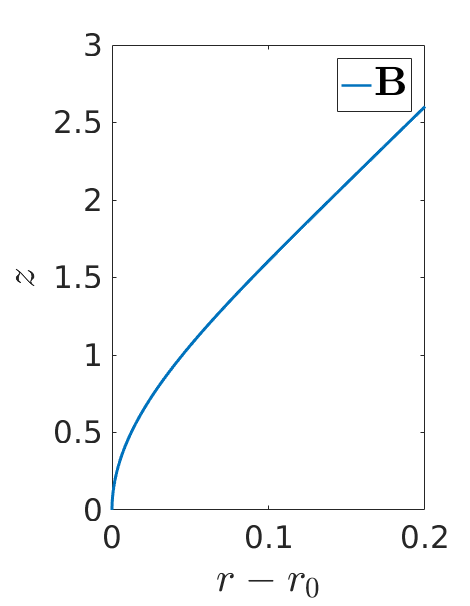}\\
	\includegraphics[width=0.3\columnwidth]{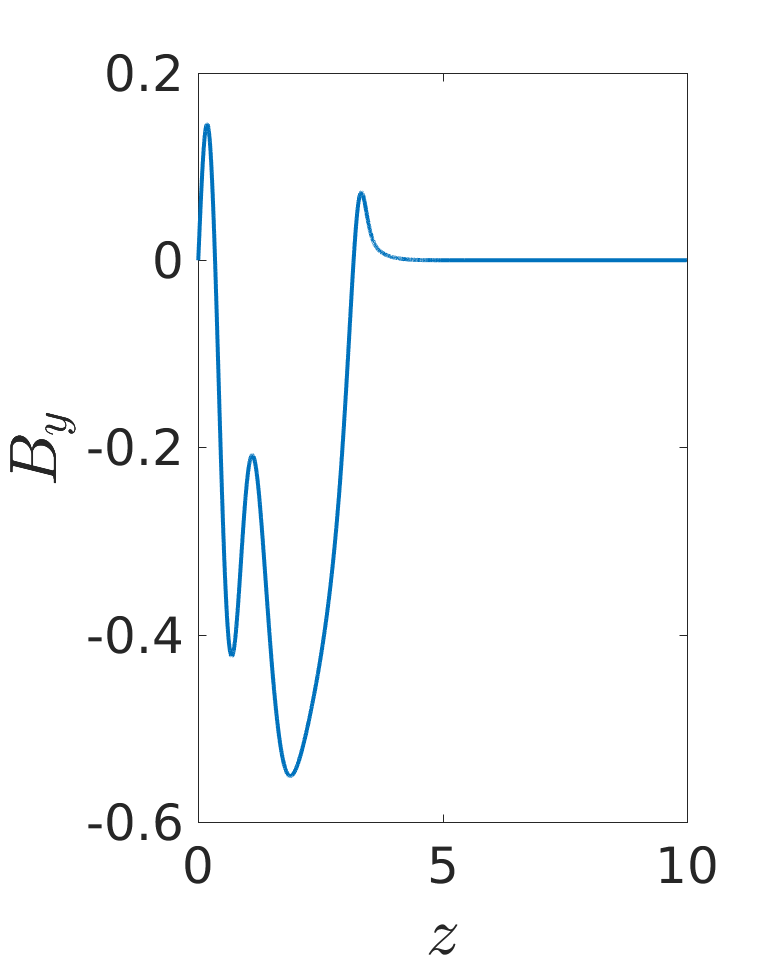}
	\includegraphics[width=0.3\columnwidth]{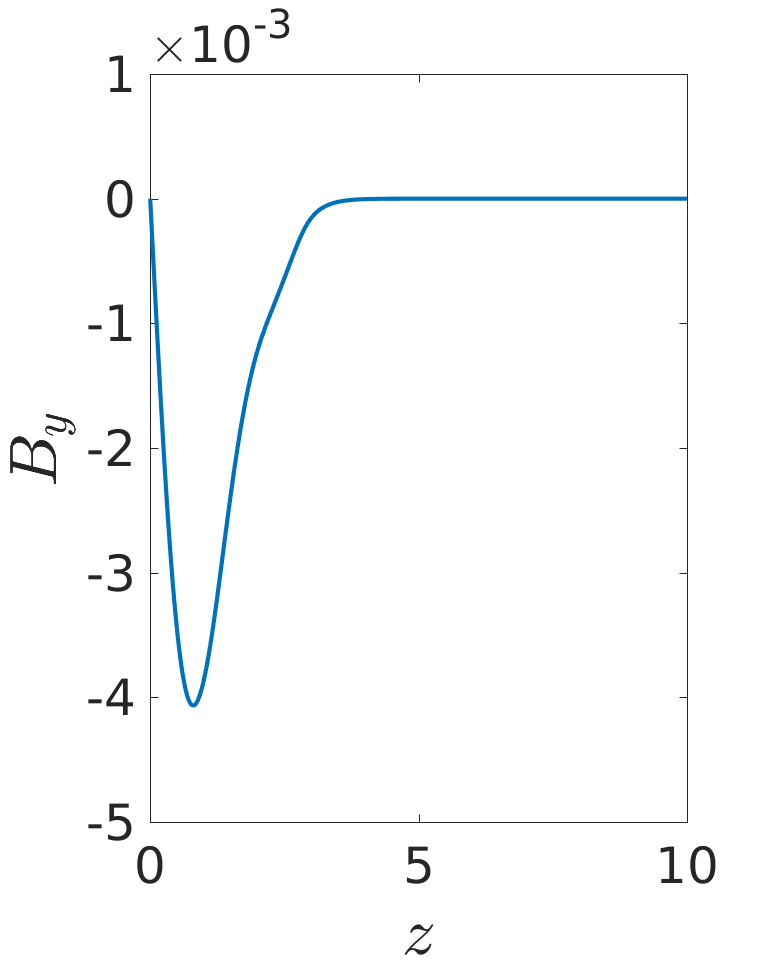}
    \end{center}
    \caption{Vertical profiles for weak field (left three) and strong field (right three) cases, $B_x(\infty)=0.1$, with midplane diffusivities of $[\eta_O,\eta_H,\eta_A]=[0.01,0.02,0.002]$ \& $[1,2,0.2]$ respectively. In our units, $B_z=1$. The top and bottom panels of each triplet plot plasma $\beta = p/p_{mag}$ and $B_y$ respectively, while the central panel plots the $\boldsymbol{B}$ field in the $rz$ plane. In all these plots, the `CstIon' diffusivity profile is used.}
    \label{fig:RLVSM.discvertstruc}
\end{figure}

We found disc vertical structure profiles to be divided into stable and unstable solutions (see Figure \ref{fig:RLVSM.discvertstruc}) when the field is aligned with the rotation. Weak field solutions with small non-ideal contributions show multiple bending of the poloidal field lines, indicating an unstable configuration due to the MRI or other instabilities \citep{Ogilvie_1998}. Strong field solutions or those with large non-ideal contributions, on the other hand, share the same general shape with a single bend and are stabilised by the strong field and/or diffusive effects present. When the field is anti-aligned with the rotation and the Hall effect is present, almost all solutions have multiple bends, and the solver often failed to converge. This is interpreted as indicating that there are no stable solutions in the anti-aligned case with the Hall effect in our 1D equilibrium model. 

Qualitatively, we find that the `Uniform' and `CstIon' diffusivity profiles give us similarly shaped vertical structures for the same set of midplane diffusivities. Differences lie in the magnitude of the horizontal $\bmB$ field, which may contrast by up to an order of magnitude (with those under `Uniform' being larger than those under `CstIon') due to the `CstIon' profile allowing $\eta_H$ and $\eta_A$ to take significantly larger values away from the midplane as $\rho$ decreases. Another difference is that the `CstIon' solutions have less smooth features in their vertical profiles at $z\sim 3-4$, when the diffusivity profile rapidly drops off to the ideal MHD regime, than their `Uniform' counterparts. However, overall, their trends in the stability of solutions and dependence on diffusivities and field strength are the same. In all the figures that follow, we plot the solutions for the more realistic `CstIon' diffusivity profile.

\subsection{Threshold magnetisation for marginal stability and variation with diffusive effects}

\begin{figure*}
    \begin{center}
	\includegraphics[width=0.45\columnwidth]{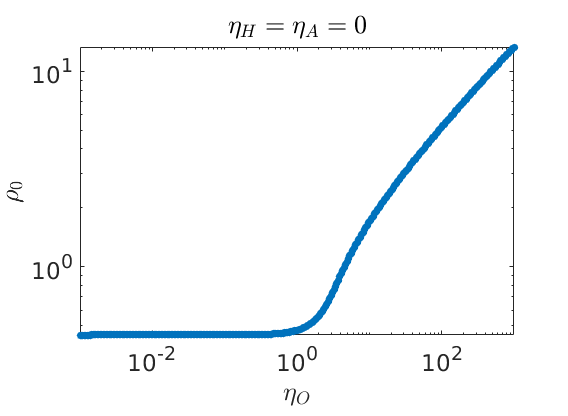}
	\includegraphics[width=0.45\columnwidth]{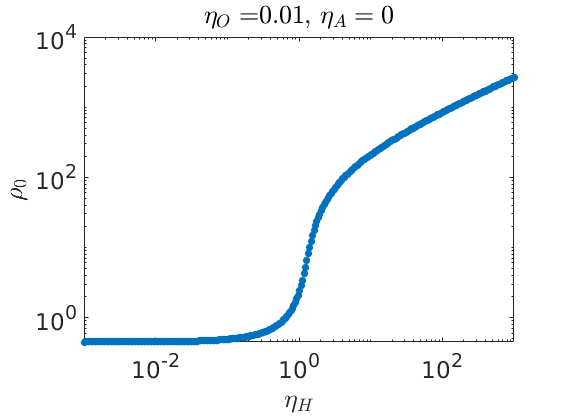}\\
	\includegraphics[width=0.45\columnwidth]{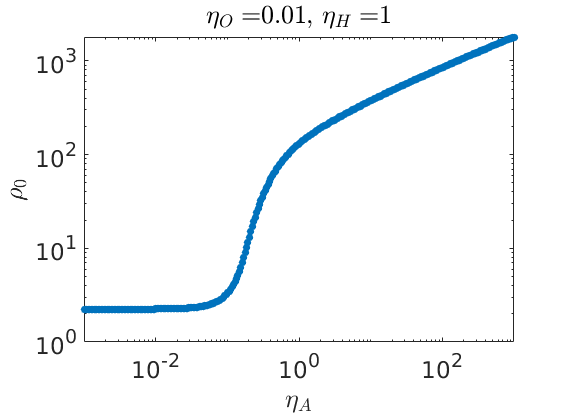}
	\includegraphics[width=0.45\columnwidth]{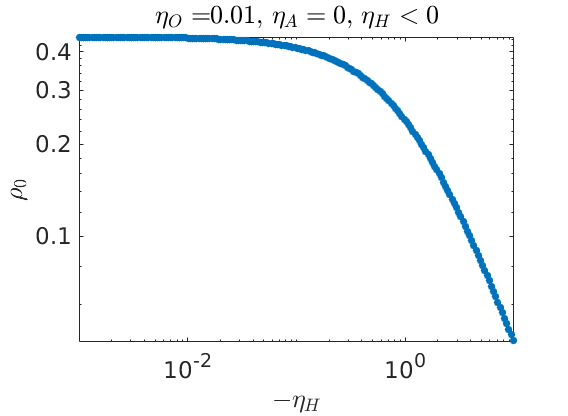}
    \end{center}
    \caption{Plots of threshold midplane densities against midplane values of the various non-ideal effects with fixed non-ideal backgrounds characterised by their midplane values. In all these plots, the `CstIon' diffusivity profile is used, while the boundary conditions are $B_x(\infty)=0.05$ and $B_y(\infty)=0$.}
    \label{fig:RLVSM.MargStab}
\end{figure*}

We conducted the marginal stability analysis described in Section \ref{sec:RLVSM.MargStab} to find the transition field strength between stable and unstable configurations. Some representative plots of our results are shown in Figure \ref{fig:RLVSM.MargStab}. As diffusive effects tend towards zero, the threshold midplane $\rho_0$ for marginal stability tends towards the value of $0.4474$, giving us a corresponding $\beta_0 \approx 0.89$. This agrees with the value found in other studies\footnote{In \citet{ParisOgilvie_2018}, a value of $B_{z'}=0.94$ is found, where $B_{z'} = B_z/\sqrt{\mu_0 \Sigma \Omega c_s }$ and $\Sigma = \int^{\infty}_\infty \rho \rmd z$. For an isothermal disc with small magnetic compression, $\rho(z)$ follows a Gaussian profile and $\rho_0/\Sigma=1/\sqrt{2\pi}$, giving us $\beta = p/p_\text{mag}\approx 0.894$. } for the largest value of $B_z$ allowing the MRI to operate in the ideal MHD regime \citep{GammieBalbus1994,ParisOgilvie_2018}. 

When $\eta_H>0$ (postive polarity configuration), all three non-ideal effects help stabilise the disc - the threshold midplane $\rho_0$ increases (i.e. critical field strength decreases) with increasing strength of each non-ideal effect. On the other hand, for negative polarity (which in our case is when $\eta_H<0$), increasing the magnitude of the Hall diffusivity decreases the threshold $\rho_0$ (critical field strength increases), showing that the Hall term has the effect of destabilising the disc\footnote{The "stability" here refers to the threshold magnetization for marginal stability, or in terms of local analysis, to the critical wavenumber for instability. However, the Hall term also has the effect of enhancing the maximum growth rate for $\eta_H (\boldsymbol{B}_z \cdot \boldsymbol{\Omega}) > 0$ (e.g., Figure 6 of \citet{WardleSalmeron_2012}), the phenomenon called the Hall-shear instability \citep{Kunz_2008}. For this reason, some authors, e.g., \citet{WardleSalmeron_2012}, describe the Hall effect for $\eta_H (\boldsymbol{B}_z \cdot \boldsymbol{\Omega}) > 0$ as "destabilizing", which is true for our disc model as well if the disc is already in the unstable regime, but irrelevant if our disc is in the regime stable to the MRI.}.

Multiple branches are observed when we varied the strength of ambipolar diffusivity under certain backgrounds. These were also observed as we varied the initial guessed value of $\rho_0$. We interpret these branches to correspond to the excitation of different modes of instability, with higher modes giving multiple bends in the disc. Marginal stability should therefore be given by the branch with the lowest $\rho_0$ values giving us the lowest order mode, where only one bend occurs.

\subsection{Flux transport rate and variation with diffusive effects}
\label{sec:RVLSM.EyTrends}


\begin{figure*}
    \begin{center}
	\includegraphics[width=0.32\columnwidth]{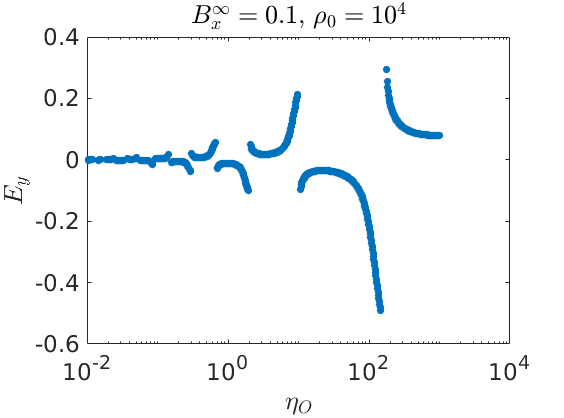}
	\includegraphics[width=0.32\columnwidth]{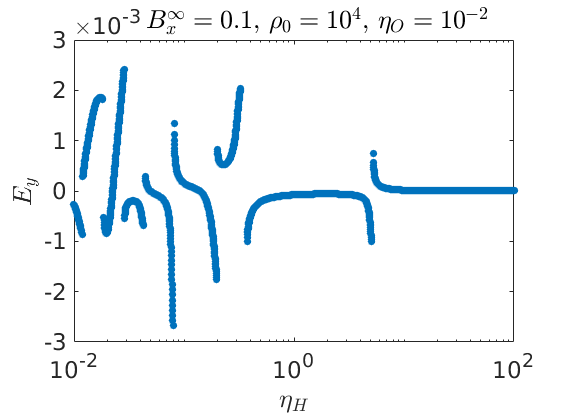}
	\includegraphics[width=0.32\columnwidth]{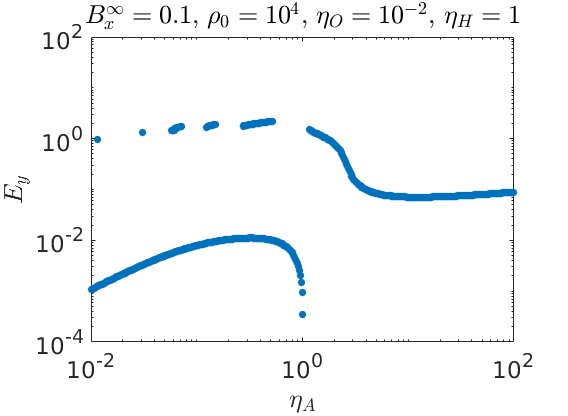}\\
	\includegraphics[width=0.32\columnwidth]{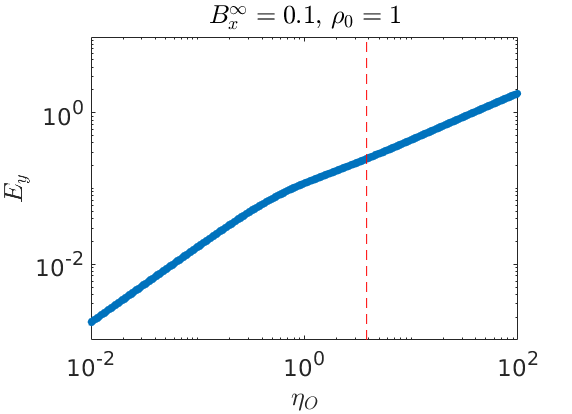}
	\includegraphics[width=0.32\columnwidth]{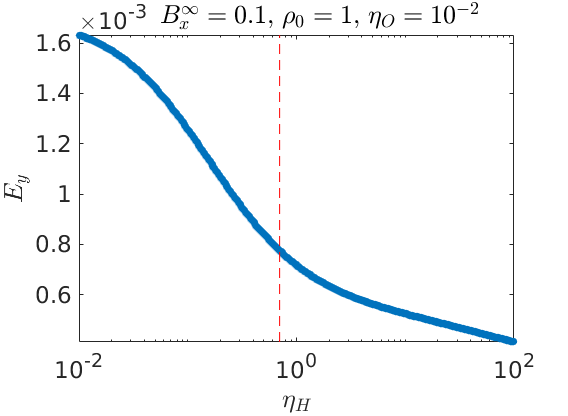}
	\includegraphics[width=0.32\columnwidth]{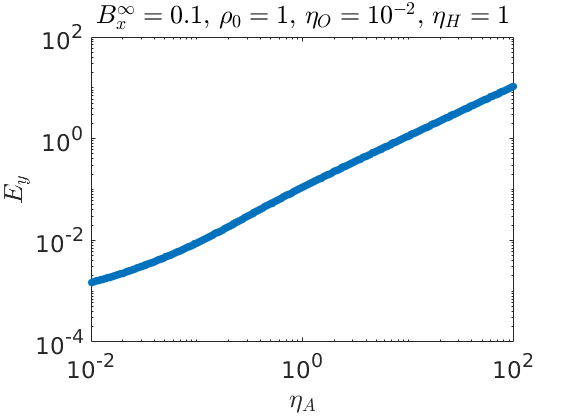}\\
	\includegraphics[width=0.32\columnwidth]{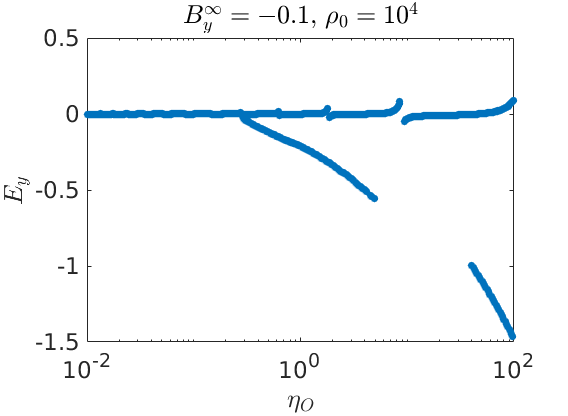}
	\includegraphics[width=0.32\columnwidth]{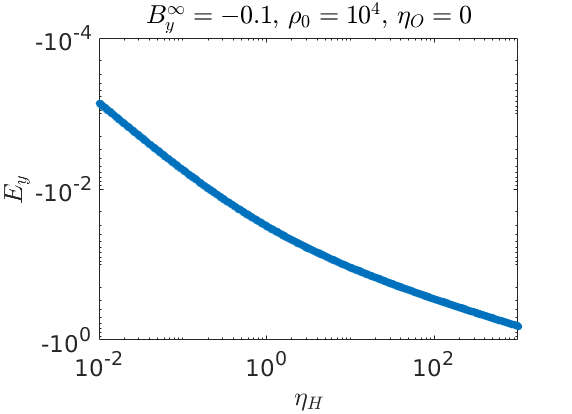}
	\includegraphics[width=0.32\columnwidth]{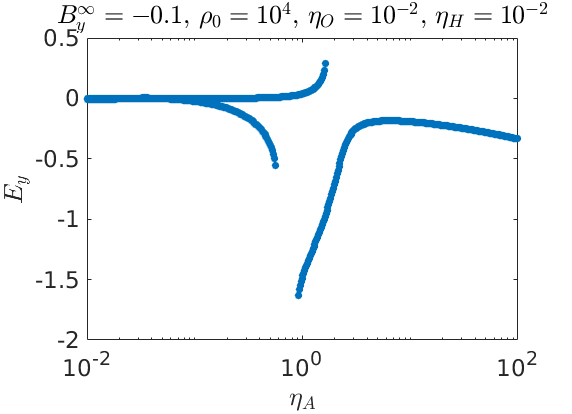}\\
	\includegraphics[width=0.32\columnwidth]{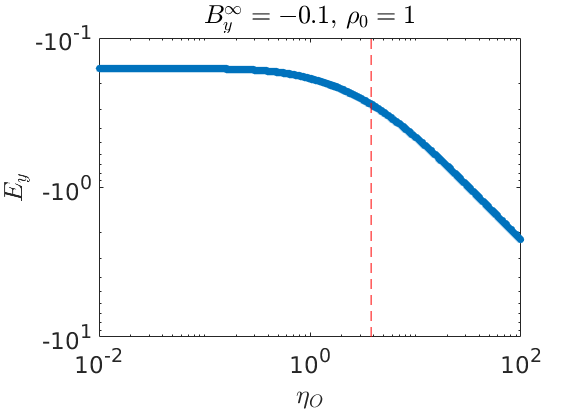}
	\includegraphics[width=0.32\columnwidth]{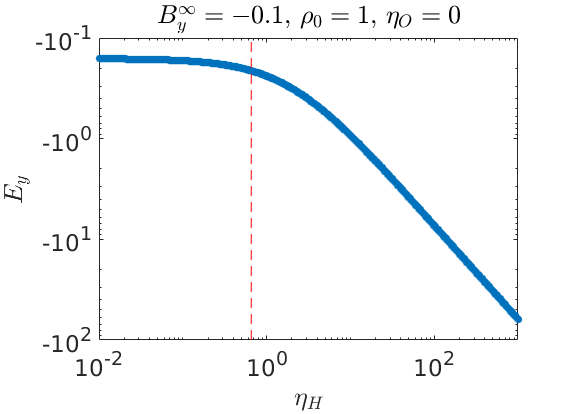}
	\includegraphics[width=0.32\columnwidth]{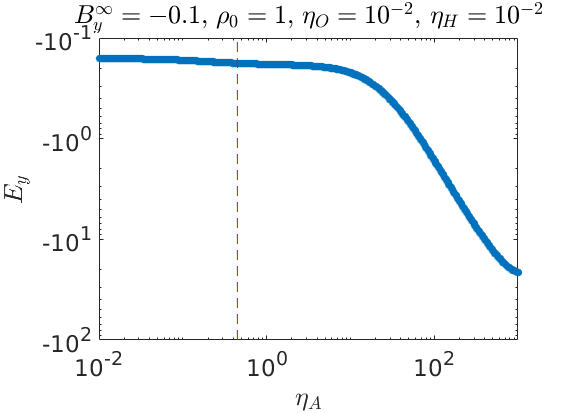}\\
	\includegraphics[width=0.32\columnwidth]{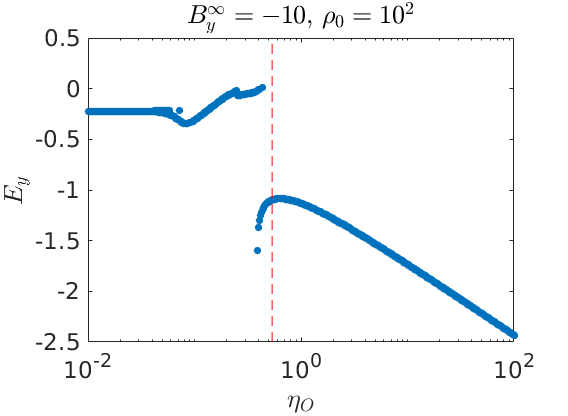}
	\includegraphics[width=0.32\columnwidth]{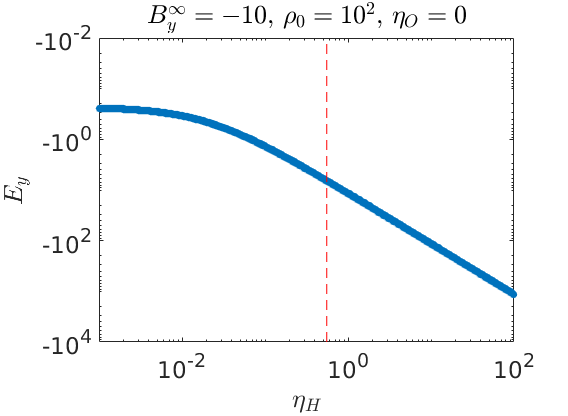}
	\includegraphics[width=0.32\columnwidth]{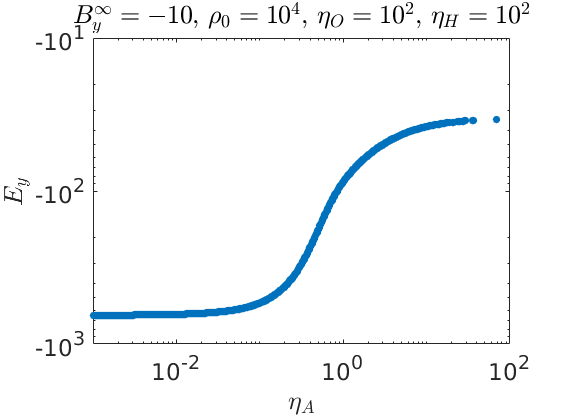}
    \end{center}
    \caption{$E_y$ and its variation with midplane diffusivities. 
    $E_y$ gives the radial flux transport velocity in units of the sound speed $c_s$. 
    The flux transport of the top two rows are inclination ($B_x(\infty)$) driven with $B_x(\infty)=0.1$, while those of the bottom three rows are outflow ($B_y(\infty)$) driven, with $B_y(\infty)=-0.1$ for rows 3 and 4, and $B_y(\infty)=-10$ for row 5. Red vertical dashed lines mark the threshold diffusivity for marginal stability calculated using the model in Section \ref{sec:RLVSM.MargStab}. Rows 1 and 3 show the weak field case ($\rho_0 = 10,000$) while rows 2 and 4 plot the strong field case ($\rho_0 =1$). Row 5 plots the case for an intermediate field strength ($\rho_0 = 100$) except the rightmost on which has a weak field ($\rho_0 = 10,000$). Background non-ideal effects are characterised by their midplane values. For rows 1 and 3, points of marginal stability are not plotted as they exist at higher diffusivity values than the range explored, whereas for the rightmost plots of both rows 2 and 5, the discs are already stable due to the background diffusivities. In all these plots, the `CstIon' diffusivity profile is used.}
    \label{fig:RLVSM.EyAlphaplots}
\end{figure*}

The variation of $E_y$ with parameters was calculated by solving for a particular set of diffusivities, then iterating to either higher/ lower diffusivities using the previous solution. This procedure allows us to identify different branches (if any) of solutions, which occur when the disc is in an unstable configuration. 

We studied most extensively the positive polarity case when $\eta_H>0$. When $\eta_H<0$ (negative polarity case), the solver failed to converge for the majority of cases. This may be indicative of the destabilising effect a negative polarity coupled with $\eta_H$ has, as found by \citet{BalbusTerquem_2001}. 

\subsubsection{Unstable solutions and bifurcations}

In all cases, weak field solutions with low background non-ideal effects show multiple branches when the direction of iteration is varied, indicative of unstable configurations. From the marginal stability analysis, we expect the point of bifurcation at the largest value of the varied diffusivity to correspond to an equivalent point there, or at least occur at a lower diffusivity than the critical one calculated. Currently our study is not fully conclusive on this, as not all branches have been exhaustively identified using our method. However, we do find that all points of bifurcation occur at lower diffusivities than the critical points calculated. On the other hand, strong field solutions with large diffusivities appear to be stable, and solutions calculated by referencing in both directions agree. 

\subsubsection{Inclination driven flux transport}

First, we explore the flux transport driven by inclination alone by setting the inclination through the value of $B_x(\infty)$. We eliminate the effect of outflow driven flux transport by setting $B_y(\infty)$ to zero.

In unstable configurations (top row of Figure \ref{fig:RLVSM.EyAlphaplots}), we find that the value of $E_y$ flips between positive and negative, with asymptotes in the diffusivity space where $E_y\to\pm\infty$. This may be indicative of collective effects such as the presence of MRI channel modes.

In stable configurations (second row of Figure \ref{fig:RLVSM.EyAlphaplots}), both Ohmic resistivity and ambipolar diffusion result in significant roughly linear increases in $E_y$, and facilitate radially outward diffusion ($E_y>0$) of the $\boldsymbol{B}$ field when $B_x(\infty)>0$ (i.e. field bends outwards) as in Figure \ref{fig:RLVSM.discvertstruc}. This agrees with the prediction from our simple analysis in Section \ref{sec:multiscale.analysis.fluxtransport} for the Ohmic case. For the ambipolar case, we expect the variation of $E_y$ with $\eta_A$ to follow the same pattern as that with $\eta_O$ as long as the `Ohm-like' term in Equation \eqref{eq:multiscale.amb.ohmlike} is significantly larger than the `Hall-like' term in Equation \eqref{eq:multiscale.amb.halllike}. This condition is indeed met for the profiles considered, as both $B_x$ and $B_y$ throughout the disc are significantly smaller than $1$, leading the disc to be largely dominated by the poloidal field component. 

The Hall effect gives no flux transport ($E_y=0$) when it is the only non-ideal effect present, but reduces the magnitude of $E_y$ when Ohmic and/or ambipolar diffusion are present, tending to a limit as $\eta_H$ becomes very large. When the Hall effect is the only non-ideal term present, $B_y=0$ throughout the disc, and there is a solution with a purely poloidal field, so that the current and Hall drift are purely in the azimuthal direction.

\subsubsection{Outflow driven flux transport}
\label{sec:RLVSM.outflow.Ey}

Next, we explore the flux transport driven by outflow alone through setting a non-zero negative value for $B_y(\infty)$. This provides a crude way of mimicking the effect of angular momentum lost vertically in an outflow. We eliminate the effect of inclination driven flux transport by setting $B_x(\infty)$ to zero. 

Again, in the unstable configurations (third row of Figure \ref{fig:RLVSM.EyAlphaplots}), we find that the value of $E_y$ flips between positive and negative, with asymptotes in the diffusivity space where $E_y\to\pm\infty$ indicative of collective effects such as the presence of MRI channel modes.

In stable configurations (bottom two rows of Figure \ref{fig:RLVSM.EyAlphaplots}), there is radially inward flux transport even when all diffusive effects disappear. This is due to flux being dragged in with the accreting gas as angular momentum is lost vertically through the disc. When $\lvert B_y(\infty) \rvert < 1$ (fourth row of Figure \ref{fig:RLVSM.EyAlphaplots}), both Ohmic resistivity and ambipolar diffusion lead to roughly linear but small increases in the magnitude of the flux transport, keeping the same radially inward direction. This is expected as the ``Ohm-like'' component of the ambipolar term would dominate since the disc field is largely poloidal in nature. The Hall effect, on the other hand, contributes significantly to radially inward flux transport when coupled with the outflow, with roughly linear increases between $\lvert E_y \rvert$ and $\eta_H$. When $\lvert B_y(\infty)\rvert > 1$ (bottom row of Figure \ref{fig:RLVSM.EyAlphaplots}), variations with Ohmic and Hall coefficients remain the same as before, but the trend with ambipolar diffusivity is reversed. Increasing the ambipolar effect now leads to a marked decrease in the inward flux transport, tending to a limiting value at high ambipolar diffusivities. This agrees with the picture described in Equation \eqref{eq:fullinduct} when the ``Hall-like'' component \eqref{eq:multiscale.amb.halllike} of the ambipolar term dominates because of the large poloidal field now present in the disc. This component operates in the opposite direction to the Hall effect in the aligned polarity configuration, hence leads to a dampening of the inward flux transport when present.

\subsection{Incompressible limit - constant density profile}
\label{sec:RLVSM.cstdensitymodel}

To help us understand and interpret the results, we examined solutions of the same set of equations in the incompressible limit for two density profiles: (i) a constant density profile with prescribed disc height, and (ii) a Gaussian density profile that takes into account the hydrostatic balance in the disc, but ignores the effect of magnetic compression. Here, we present the results of (i), which admit approximate analytic solutions, and how they inform the relations between flux transport, diffusive effects and inclination/outflow. 

\subsubsection{Approximate analytic solutions}

We solve the same system of Equations \eqref{RLVSM1.1}$-$\eqref{RLVSM1.5} and set $\rho = \cst$. The permeability of free space, $\mu_0$ is set to 1. We assume a Keplerian disc and set the orbital parameter $q = -\p \ln{\Omega}/\p\ln{r}$ to $3/2$. For simplicity and the possibility of admitting analytic solutions, we also assume that $\eta_O= \cst$, and constant ionisation fractions in the disc. This means $\Tilde{\eta}_H,\Tilde{\eta}_A=\cst$, where they are field-independent coefficients given by
\begin{equation}
    \Tilde{\eta}_H = \eta_H/ \lvert \bmB \rvert,
\end{equation}
\begin{equation}
    \Tilde{\eta}_A = \eta_A/ \lvert \bmB \rvert ^ 2.
\end{equation}
This leads to the following form of the modified Ohm's law including the non-ideal effects:
\begin{equation}\label{E_x}
    E_x = - v_y B_z + \eta_O J_x 
    + \Tilde{\eta}_H J_y B_z
    + \Tilde{\eta}_A [ J_x B_z^2 + (J_x B_y - J_y B_x) B_y ],
\end{equation}
\begin{equation}\label{E_y}
    E_y = v_x B_z + \eta_O J_y 
    - \Tilde{\eta}_H J_x B_z 
    + \Tilde{\eta}_A [ J_y B_z^2 - (J_x B_y - J_y B_x) B_x ],
\end{equation}
with
\begin{equation}\label{Amp_x}
    J_x = - \f{\rmd B_y}{\rmd z},
\end{equation}
\begin{equation}\label{Amp_y}
    J_y = \f{\rmd B_x}{\rmd z}.
\end{equation}

The ODE system is linear in the Ohmic and Hall only cases, and can be reduced to give us the following wave equation:
\begin{equation}\label{B_xWaveEq}
    \f{\rmd^2 B_x}{\rmd z^2} + k^2 B_x = 0,
\end{equation}
with $k^2$ given by
\begin{equation} \label{knumberequation}
    k^2 = 
    \f{3 \Omega^2}{v_{az}^2}
    \left[ 1 + \f{\eta_O^2 \Omega^2}{v_{az}^4}
    \left( \f{1}{1 + \Tilde{\eta}_H B_z \Omega/ 2 v_{az}^2 }
    \right)
    + \f{2 \Tilde{\eta}_H B_z \Omega}{v_{az}^2}
    \right]^{-1},
\end{equation}
where $v_{az}= B_z/\sqrt{\rho}$ is the vertical component of the Alfv\'en velocity.

If we expect the disc to satisfy the symmetry and boundary conditions described in Section \ref{sec:shooting}, then the relevant solution is
\begin{equation}
    B_x = B_x(\infty) \f{\sin{kz}}{\sin{kH}},
\end{equation}
where the surfaces of the disc are at $z = \pm H$, and above and below the disc we have $B_x = \pm B_x(\infty)$ respectively.

The other variables can be deduced from $B_x$ and the bonudary conditions. If $B_y = \pm B_y(\infty)$ at the upper and lower boundaries (where $B_y(\infty)$ can be zero), we have
\begin{equation}
\begin{aligned}
    B_y =~& \f{ \Omega \eta_O}{2v_{az}^2}
    \left( \f{1}{1 + \Omega \Tilde{\eta}_H B_z/2 v_{az}^2} \right)
    B_x(\infty)
    \left( \f{z}{H} - \f{\sin{kz}}{\sin{kH}} \right) \\
    & + B_y(\infty) \f{z}{H}.
\end{aligned}
\end{equation}

The ambipolar contribution, on the other hand, has both a linear and a nonlinear part. The linear part behaves as an additional Ohmic resistivity. The nonlinear part is of order $B_{\text{horizontal}}^2$, which is small if the surface value of $B_\text{horizontal}$ is also small. We limit our analysis here to the cases when $B_y(\infty)<1$, and therefore only the ``Ohm-like'' component of the ambipolar term contributes significantly to the flux transport. The system can hence be roughly solved as the same set of linear equations with a modified Ohmic resistivity $\eta_{O,\text{mod}} = \eta_{O} + \Tilde{\eta}_{A} B_z^2$ when ambipolar diffusion is also included.

\subsubsection{Relation to \citet{BalbusTerquem_2001}}

Equation \eqref{knumberequation}, which gives us the wavenumber of the field solution, is mathematically identical to the wavenumber formula (Equation (46)) of a marginally stable Ohm and Hall only MRI mode derived in \citet{BalbusTerquem_2001}. The problem they examined, however, is different from the one investigated here. They determined the local stability of a differentially rotating disc threaded by a weak vertical field and were looking at plane wave disturbances of the form $\exp{(i k z - i \omega t)}$. The Boussinesq limit was used as it corresponds to fluid displacements within a local approximation in $z$, rendering vertical structure unimportant. Our work, on the other hand, solves for the disc vertical structure, but deliberately makes the physically unrealistic uniform density approximation to simplify the equations. Our solution therefore must obey the boundary conditions determined by field inclination and symmetry criteria, which are not required in the problem examined by \citet{BalbusTerquem_2001}.  

\subsubsection{Marginal stability and variation with magnetisation and diffusivities}

As in Section \ref{sec:RLVSM.MargStab}, we examine the case of a disc at marginal stability to the MRI. In our approximate analytic model, because the equations are already linear, we solve the same set of equations for the perturbed fluid variables. As before, these perturbations have opposite symmetry about the midplane to the fluid variables, which in our case means that $\delta B_x \propto \cos{kz}$. The boundary condition is now $\delta B_x = 0$ and $\delta B_y = 0$, to meet the fixed inclination and outflow conditions. We expect the marginally stable mode to allow just one bending of the field through the disc, hence we require the corresponding critical wavenumber to satisfy $k_\text{crit}H = \pi /2 $. The condition for marginal stability then becomes:

\begin{equation}
    2 q \Omega^2 = 
    \f{\pi^2 v_{az}^2}{4 H^2}
    \left[ 1 + \f{\eta_O^2 \Omega^2}{v_{az}^4}
    \left( \f{1}{1 + \Tilde{\eta}_H B_z \Omega/ 2 v_{az}^2 }
    \right)
    + \f{2 \Tilde{\eta}_H B_z \Omega}{v_{az}^2}
    \right].
\end{equation}
We can recast this in terms of a dimensionless magnetisation parameter:
\begin{equation}
    \mu = \f{B_z}{\sqrt{\Sigma H \Omega^2}},
\end{equation}
where $\Sigma = 2 \rho H$ is the disc surface density, giving us
\begin{equation}\label{eq:IncomModel.mageq}
    \f{4 q}{\pi^2} = 
    \mu^2 + 
    \left(\f{\eta_O }{2 H^2 \Omega }\right)^2
    \left( \f{1}{1 + \Tilde{\eta}_H \sqrt{\rho} / 2\sqrt{2} H \mu }
    \right) \f{1}{\mu^2}
    + \f{\sqrt{2 \rho} \Omega}{\Tilde{\eta}_H} \mu.
\end{equation}

Magnetisations that give us a stable solution can be determined when the left hand side of Equation \eqref{eq:IncomModel.mageq} is larger than the right hand side. We rearrange the terms in the equation to give us the following quartic, with the inequality giving us the condition for a stable solution as:
\begin{equation}
\begin{aligned}
    & f(\mu) = \mu^4 + 5\mathcal{H}\mu^3    
    + (4 \mathcal{H}^2 - \mathcal{Q})\mu^2
    - \mathcal{Q} \mathcal{H} \mu 
    + \mathcal{R}^2 
    \geq 0,
\end{aligned}
\end{equation}
where we have defined the following dimensionless parameters:
\begin{equation}
    \mathcal{R} = \f{\eta_O }{2 H^2 \Omega},
\end{equation}
\begin{equation}
    \mathcal{H} = \f{\Tilde{\eta}_H }{2 H } \sqrt{\f{\rho}{2}}
    = \f{\Tilde{\eta}_H }{2 \sqrt{2} H} \f{\sqrt{\rho}}{B_z}
    = \f{\eta_H }{2 \sqrt{2} H} \f{1}{v_{az}},
\end{equation}
\begin{equation}
    \mathcal{Q} = \f{4q}{\pi^2}.
\end{equation}

\begin{figure}
    \begin{center}
	\includegraphics[width=0.5\columnwidth]{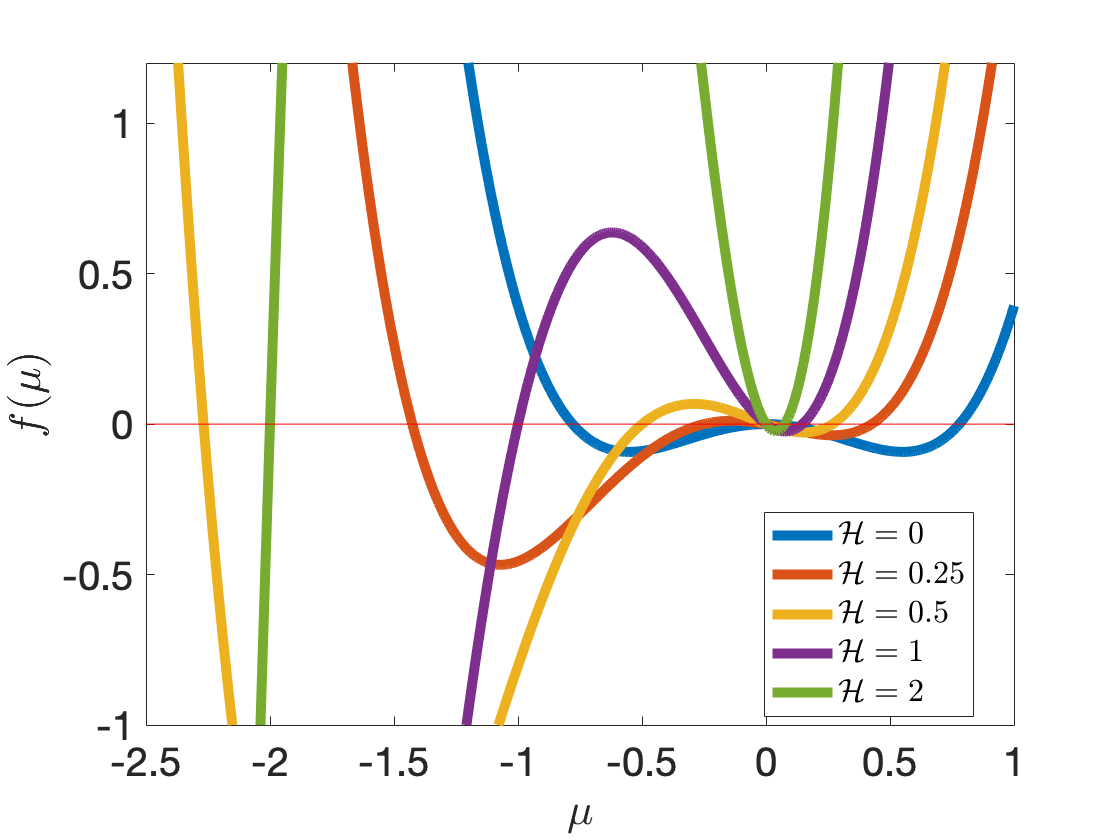}
    \end{center}
    \caption{Plot of $f(\mu)$ vs. $\mu$ for various Hall parameters $\mathcal{H}$, with $\mathcal{R}=0$. Increasing $\mathcal{R}$ does not alter the shape of the plot but shifts $f$ up by $\mathcal{R}^2$. As $\mathcal{H}$ increases, the unstable ($f<0$) region expands when $\mu<0$, while it contracts when $\mu>0$. $f\to\infty$ as $\mu\to\pm\infty$ for all values of $\mathcal{H}$. The red dotted line marks $f=0$, the boundary between stable and unstable disc configurations. Solutions below the line are unstable, while solutions above the line are stable.}
    \label{fig:IncomModel.magnetisation_plot}
\end{figure}

The shape of $f(\mu)$ vs. $\mu$ (see Figure \ref{fig:IncomModel.magnetisation_plot}) is determined by the Hall diffusivity only, while Ohmic diffusivity stabilises the entire profile by adding a constant term to $f(\mu)$. At high magnetisations, the disc is always stabilised whether the field is aligned with the rotation or not. In the intermediate region, a large Hall parameter extends the region for instability when the polarity is negative, but has a stabilising effect when the polarity is positive. This agrees with our previous result that the Hall effect is responsible for destabilising the disc in the anti-aligned case, while it is stabilising in the aligned case.


\subsubsection{Flux transport}

For flux transport, the constant density model gives us this simple relation:
\begin{equation}\label{eq:RLVSM.Ey.analytic}
\begin{aligned}
    E_y = ~& \left( \f{2 B_z^2}{\Omega\rho}
    + \Tilde{\eta}_H B_z \right) \f{\rmd B_y}{\rmd z}
    + \eta_O \f{\rmd B_x}{\rmd z}\\
    = ~ & \f{\eta_O}{H} B_x(\infty)
    + \left( \f{2B_z^2}{\Omega\rho}
    + \eta_H \right) \f{B_y(\infty)}{H},
\end{aligned}
\end{equation}
with the radial flux transport velocity given by $E_y /B_z$. 

We can see that flux transport is directly proportional to the inclination and Ohmic diffusivity, which is also the case in the numerical solutions of the compressible model. When no diffusive effects are present, the relation suggests that the presence of an outflow would still drive inward flux accretion, with a large magnetisation (small $\rho$) giving the fastest rate. This is indeed what we observed in the compressible model, and can be interpreted to be a result of the loss of angular momentum driving inward gas accretion, which in turn advects the flux frozen into the gas along with it. A larger magnetisation would mean a higher degree of flux freezing, hence advection with the accretion flow.

The flux transport rate modification due to the Hall effect coupled with inclination is not explained here, but the constant density model predicts that the Hall effect coupled with a non-zero $B_y$ at infinity would also lead to a non-zero flux transport. A positve $\eta_H$ coupled with the negative $B_y^+$ expected for a wind would therefore lead to a negative $E_y$ proportional to $\eta_H$ and $B_y^+$, signifying a radially inward accretion of flux. This is indeed what we find in the solutions of Section \ref{sec:RLVSM.outflow.Ey}. Such an effect has been noted before in passing in Hall-wind shearing box simulations \citep{Bai_2014} and also in the global simulations of \citet{BaiStone2017}. It is also noteworthy that the $\eta_H$ term in our flux transport equation is essentially of the same form as Equation 10 of \citep{BaiStone2017}, although our analytic model is more rigorous and would therefore provide a more accurate estimate.

\subsection{Incompressible limit - Gaussian density profile}

For our second incompressible limit, we imposed a Gaussian density profile that takes into account the balance between pressure and gravity in an isothermal disc, but ignores the effect of magnetic compression from the horizontal field components. Solutions are calculated numerically using the same method as the compressible case, but with the density profile fixed. Comparing the solutions of this model to that of the compressible model allows us to determine whether magnetic compression has a significant effect on the flux transport observed.

The Gaussian density profile imposed is the following:
\begin{equation}
    \rho = \rho_0 \exp{\left( - z^2 / 2 \right)},
\end{equation}
where $\rho_0$ is the midplane density value. 

We found that solutions from the Gaussian incompressible model are very similar both in form and magnitude to those of the compressible model. The only differences arise where bifurcations in the $E_y$-diffusivity space occur in the compressible model. There, the Gaussian incompressible model yields vertical asymptotes on either side of the point of divergence instead. This can be explained by the fact that as the instability encounters a point of resonance, magnetic compression can provide the required change in density structure to dampen its effect, hence allowing the solution branches to diverge rather than tend towards a vertical asymptote. Therefore we can conclude that magnetic compression has no significant effect on the flux transport, in the parameter space we have explored.   


\section{The Guilet \& Ogilvie approach}
\label{sec:GuiletOgilvie}

Our second semi-analytic model follows the scalings used in \citet{GuiletOgilvie2012} (hereafter GO1). This scheme considers the case when the vertical field $B_z$ dominates over $B_r$ and $B_\phi$ by a factor of $\epsilon^{-1}$. The motivation for this is to put the effects from large scale radial gradients (the $\p_r$ terms) on the same footing as inclination and outflow. Again, instead of using the residual velocity $\bmv = \bmu - r \Omega \hat{\bme}_\phi$, we use the full velocity $\bmu$ (we do not relabel it as $\bmv$ here). Compared to the multiscale approach, we make the following adjustments:
\begin{equation}\label{eq:GOTermsZero}
    B_{r0}, B_{\phi0}, B_{z1}, u_{r0}, u_{\phi0}, u_{z0}, u_{z1} = 0,
\end{equation}
The next order terms are assumed to be dominant, and the higher order terms are neglected.
We extend the analysis of GO1 to derive the case for a laminar disc with all non-ideal diffusivities present. 

\subsection{Leading order equations}

The leading order radial component of the equation of motion is
\begin{equation}
    -\rho_0 r \Omega^2 = -\rho_0 \p_r \Phi_m,
\end{equation}
which describes the centrifugal force balance against the inward gravitational pull. For a Keplerian disc this gives us $\Omega_0 = (GM/r^3)^{1/2}$.

The vertical component of the equation of motion at leading order gives us the hydrostatic equilibrium between the vertical gravitational force and the vertical pressure gradient:
\begin{equation}
    0 = -\rho_0 \Psi \zeta - \p_\zeta p_0.
\end{equation}
For an isothermal disc, we have the familiar solution
\begin{equation}
    \rho_0 = \f{\Sigma_0}{\sqrt{2\pi} H_0}\exp{\left(-\f{\zeta^2}{2H_0^2}\right)},
\end{equation}
where $H_0 = c_s/\Omega_0$ is the isothermal scaleheight and $\Sigma_0(r,\tau)$ is the surface density.

The horizontal components of the equation of motion and the induction equation come at the next order, due to the terms that we set to zero in Equation \eqref{eq:GOTermsZero}. They are identical to the equations in GO1 except for the addition of the Hall and ambipolar terms:
\begin{equation}\label{eq:GOdiff1.1}
\begin{aligned}
    - 2\rho_0 \Omega_0 u_{\phi1} = &
    - \rho_0 \p_r \Psi \f{1}{2} \zeta^2 - \p_r \left( p_0 + \f{B_{z0}^2}{2 \mu_0} \right)
    + \f{B_{z0}}{\mu_0} \p_\zeta B_{r1},
\end{aligned}
\end{equation}
\begin{equation}\label{eq:GOdiff1.2}
\begin{aligned}
    \rho_0 u_{r1} \f{1}{r} \p_r ( r^2 \Omega_0 ) = &
    \f{B_{z0}}{\mu_0} \p_\zeta B_{\phi1} ,
\end{aligned}
\end{equation}
\begin{equation}\label{eq:GOdiff1.3}
\begin{aligned}
    0 = B_{z0} \p_\zeta u_{r1} 
    + \p_\zeta[& (\eta_{O0}+\eta_{A0}) \left(\p_\zeta B_{r1}
  - \p_r B_{z0}\right) + \eta_{H0}
  \p_\zeta B_{\phi1}  ],
\end{aligned}
\end{equation}
\begin{equation}\label{eq:GOdiff1.4}
\begin{aligned}
    0
    = & B_{z0} \p_\zeta u_{\phi1}+B_{r1} r \p_r \Omega_0 \\
    & + \p_\zeta\left[ (\eta_{O0}   +  \eta_{A0})
  \p_\zeta B_{\phi1}  
  -  \eta_{H0}  \left(\p_\zeta B_{r1}
  - \p_r B_{z0} \right)
    \right].
\end{aligned}
\end{equation}
Under these scalings, we can see that Ohmic and ambipolar diffusivities have the same effect on the disc dynamics. The Hall term comes in at $\pi/2$ phase difference with the Ohmic and ambipolar terms in the induction equation. This reflects the nature of the Hall term being the magnetic field cross multiplied once with the Ohmic term, while the ambipolar term is cross multiplied twice with the Ohmic term hence acts in the same direction at leading order.

This analysis yields four linear equations for the unknowns $u_{r1}$, $u_{\phi1}$, $B_{r1}$ and $B_{\phi1}$. The linearity is a result of the assumption that we are examining the case of small deviations from orbital motion and a vertical magnetic field. 

\subsection{Non-dimensionalisation}

We follow the same approach as GO1 in non-dimensionalising our equations. We relegate the definitions to Appendix C (see online supplementary materials), but note that $\Tilde{\rho}, u_r, u_\phi, b_r, b_\phi, \Tilde{\eta}_{O}, \Tilde{\eta}_{H}, \Tilde{\eta}_A, \zeta$ are the non-dimensionalised forms of  $\rho, u_{r1}, u_{\phi1}, B_{r1}, B_{\phi1}, {\eta}_{O0}, {\eta}_{H0}, {\eta}_{A0}, \zeta$ respectively. 

We assume a point-mass potential and circular Keplerian orbital motion at leading order. An isothermal equation of state is also employed for simplicity.

The density profile in dimensionless form is
\begin{equation}
    \Tilde{\rho} = \f{1}{\sqrt{2\pi}} \exp{\left(-\zeta^2/2\right)},
\end{equation}
and the differential equations \eqref{eq:GOdiff1.1}$-$\eqref{eq:GOdiff1.4} are  rewritten as
\begin{equation}\label{eq:GODiffEqns2.1}
\begin{aligned}
    & - 2 u_\phi - \f{1}{\beta_0 \Tilde{\rho}} \p_\zeta b_r
    \\
    & \qquad \qquad\qquad
    =
    \f{3}{2} + D_H - D_{\nu\Sigma}
    + \left( \f{3}{2} -  D_H \right) \zeta^2 
    - \f{ D_B}{\beta_0 \Tilde{\rho} } ,
\end{aligned}
\end{equation}
\begin{equation}\label{eq:GODiffEqns2.2}
\begin{aligned}
    & \f{1}{2} u_r - \f{1}{\beta_0 \Tilde{\rho}} \p_\zeta b_\phi
    = 0,
\end{aligned}
\end{equation}
\begin{equation}\label{eq:GODiffEqns2.3}
\begin{aligned}
    & - \p_\zeta\left( \left[ \Tilde{\eta}_{O}+\Tilde{\eta}_A \right] \p_\zeta  b_r 
    + \Tilde{\eta}_H \p_\zeta b_\phi \right)
    - \p_\zeta u_r \\
    & \qquad\qquad\qquad
    = - D_B \p_\zeta \left( \Tilde{\eta}_{O}+\Tilde{\eta}_A \right) ,
\end{aligned}
\end{equation}
\begin{equation}\label{eq:GODiffEqns2.4}
\begin{aligned}
    & - \p_\zeta\left( - \Tilde{\eta}_H \p_\zeta  b_r 
    + \left[ \Tilde{\eta}_{O}+\Tilde{\eta}_A \right] \p_\zeta b_\phi \right)
    - \p_\zeta u_\phi + \f{3}{2} b_r \\
    & \qquad\qquad\qquad
    = D_B \p_\zeta \Tilde{\eta}_{H} ,
\end{aligned}
\end{equation}
where 
\begin{equation}
    \beta_0 \equiv \f{\mu_0}{B_z^2}\f{\Sigma c_s^2}{H},
\end{equation}
\begin{equation}
    D_H \equiv \f{\p \ln{H}}{\p \ln{r}},
\end{equation}
\begin{equation}
    D_{\nu\Sigma} = 2D_H - \f{3}{2} + \f{\p \ln{\Sigma}}{\p \ln{r}},
\end{equation}
\begin{equation}
    D_B \equiv \f{\p \ln{B_z}}{\p \ln{r}}.
\end{equation}
A key thing to note here is that the LHS of Equations \eqref{eq:GODiffEqns2.1}-\eqref{eq:GODiffEqns2.4} forms the differential system, while the RHS are source terms that drive the advection and diffusion of flux. These equations are in many ways similar to those we obtained in Section \ref{sec:RLVSM.model}. The first two terms on the LHS of Equation \eqref{eq:GODiffEqns2.1} correspond to the two terms in Equation \eqref{RLVSM1.1}, with the only difference being the addition of large scale radial gradient and Keplerian source terms on the RHS. Equation \eqref{eq:GODiffEqns2.2} is identical to Equation \eqref{RLVSM1.2}, while Equations \eqref{eq:GODiffEqns2.3} and \eqref{eq:GODiffEqns2.4} are linearised versions of Equations \eqref{RLVSM1.4} and \eqref{RLVSM1.5} respectively.

\subsection{Boundary conditions}

We use the same boundary conditions as determined in Section 3.2 of GO1, which come from analysing the same expected symmetry of the solutions as Section \ref{sec:shooting}, with $v_{r1}$ and $v_{\phi1}$ being even functions of $\zeta$, while $B_{r1}$ and $B_{\phi1}$ are odd. The following quantities can then be determined to vanish exponentially fast as $\zeta\to\pm\infty$:
\begin{equation}
    \rho u_r \to 0,
\end{equation}
\begin{equation}
    \rho u_\phi \to 0,
\end{equation}
\begin{equation}\label{eq:GO.brInftybonds}
    b_r - (D_B\zeta \pm b_{rs}) \to 0,
\end{equation}
\begin{equation}\label{eq:GO.bphiInftybonds}
    b_\phi - (\pm b_{\phi s}) \to 0.
\end{equation}
The symmetry of the solutions also constrain the following mid-plane ($\zeta=0$) values to be the following:
\begin{equation}
    \p_\zeta u_r = 0,
\end{equation}
\begin{equation}
\p_\zeta u_\phi = 0,
\end{equation}
\begin{equation}
b_r = 0,
\end{equation}
\begin{equation}
b_\phi = 0.
\end{equation}
A non-vanishing $b_{\phi s}$, the value of $b_\phi$ at $\zeta\to\infty$, is again included for mimicking the effect of angular momentum removal by a MCW or magnetic braking due to interaction with an external medium. 
As noted in GO1, the boundary conditions are homogeneous, except for the linear source terms proportional to $D_B$, $b_{rs}$ and $b_{\phi s}$ in Equations \eqref{eq:GO.brInftybonds} and \eqref{eq:GO.bphiInftybonds}.

\subsection{Form of the solution for a laminar inviscid disc}

The linearity of the equations means that the general solution is a linear combination of the solution vectors corresponding to each source term, which appear either on the right hand side of equations \eqref{eq:GODiffEqns2.1}$-$\eqref{eq:GODiffEqns2.4}, or as a non-vanishing boundary condition at infinity in the form of $b_{rs}$ and $b_{\phi s}$. We can thus write the general solution $\boldsymbol{X}=\{u_r,u_\phi,b_r,b_\phi\}$ as:
\begin{equation}\label{eq:GO.formsol}
\begin{aligned}
    \boldsymbol{X} =& \boldsymbol{X}_\mathrm{K}
    + \boldsymbol{X}_\mathrm{DH}
    + \boldsymbol{X}_\mathrm{D\nu\Sigma} D_\mathrm{\nu\Sigma}
    + \boldsymbol{X}_\mathrm{DB} D_\mathrm{B}\\
    & \qquad
    + \boldsymbol{X}_\mathrm{brs} b_{rs}
    + \boldsymbol{X}_\mathrm{b\phi s} b_{\phi s},
\end{aligned}
\end{equation}
where $\boldsymbol{X}_\mathrm{DH}$ is the solution vector corresponding to the source term proportional to $D_H$ and so on. $\boldsymbol{X}_K$ corresponds to the solution vector when $D_H,D_{\nu\Sigma},D_B,b_{rs},b_{\phi s}=0$, where the source terms arise only from the radial derivatives of the leading order Keplerian, gravitational and geometric terms. Following GO1, we also define
\begin{equation}
    \boldsymbol{X}_\mathrm{hyd} = \boldsymbol{X}_\mathrm{K} 
    + \boldsymbol{X}_\mathrm{DH},
\end{equation}
as the solution with hydrodynamic (`Hydro') source terms $D_H=1$ (corresponding to a disc with a constant aspect ratio $H/r$) and $D_{\nu\Sigma}=0$ (for a steady accretion flow far from the inner boundary). Under these definitions, we have
\begin{equation}
    \boldsymbol{X} = \boldsymbol{X}_\mathrm{hyd}
    + \boldsymbol{X}_\mathrm{D\nu\Sigma} D_\mathrm{\nu\Sigma}
    + \boldsymbol{X}_\mathrm{DB} D_\mathrm{B}
    + \boldsymbol{X}_\mathrm{brs} b_{rs}
    + \boldsymbol{X}_\mathrm{b\phi s} b_{\phi s}.
\end{equation}

The solution depends in a non-linear way only on the parameters $\beta_0$, $\eta_O$, $\eta_H$ and $\eta_A$ (we drop the $\Tilde{}$ in the text from this point onwards but refer to the non-dimensionalised diffusivities). Since we consider a laminar inviscid disc, we set $\alpha$ to zero, and can neglect the $\mathcal{P}$ dependence. $\eta_O$ and $\eta_A$ have the same effect at the order we are considering, so we only need to examine one of them, which we choose to be $\eta_O$. For each triplet of values of the three parameters $\beta_0$, $\eta_O$ and $\eta_H$, one needs to compute the six solution vectors for each of the terms on the right hand side of Equation \eqref{eq:GO.formsol}. The general solution is then given by a linear combination of these solution vectors with the appropriate coefficients. 

\subsection{Numerical method}

To solve the system of ODEs, we employ the same pseudo-spectral method of GO1 using a decomposition on a basis of Whittaker cardinal functions. These functions are well suited to problems on an infinite interval \citep{Boyd_2001,Latter_etal_2010}. The use of Whittaker cardinal functions implicitly imposes the condition that the variables have to vanish exponentially fast at infinity. Following GO1, we replace $b_r$ and $b_\phi$ with the following variables:
\begin{equation}
    \Tilde{b}_r \equiv b_r - D_B \zeta - b_{rs} \tanh{(\zeta^3)},
\end{equation}
\begin{equation}
    \Tilde{b}_\phi \equiv b_\phi - b_{\phi s} \tanh{(\zeta^3)}.
\end{equation}
to satisfy this condition. We can then see that these new variables do vanish exponentially fast at infinity from the boundary conditions given in equations \eqref{eq:GO.brInftybonds} and \eqref{eq:GO.bphiInftybonds}.

\subsection{Flux transport for uniform diffusivities}


\begin{figure*}
    \begin{center}
	\includegraphics[width=0.33\columnwidth]{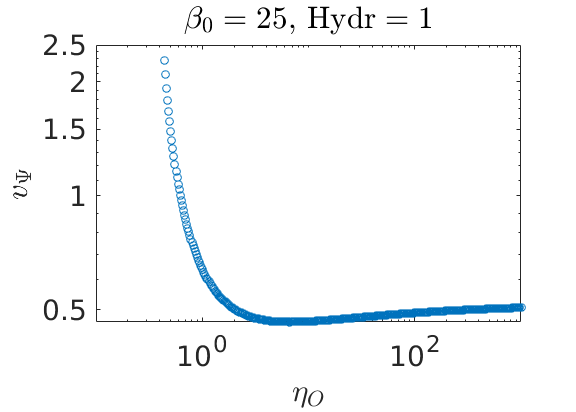}
	\includegraphics[width=0.33\columnwidth]{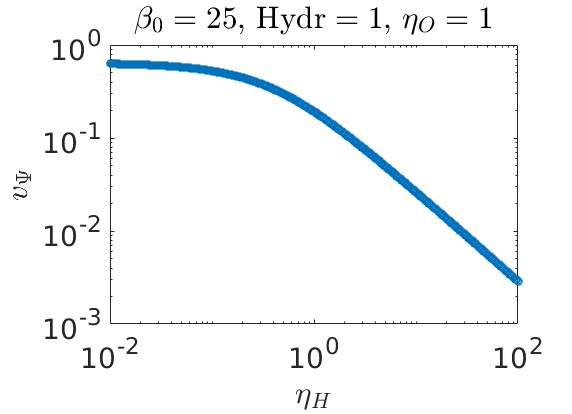}
	\includegraphics[width=0.33\columnwidth]{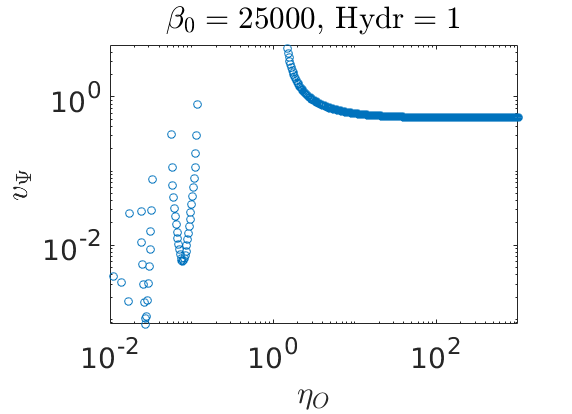}\\
	\includegraphics[width=0.33\columnwidth]{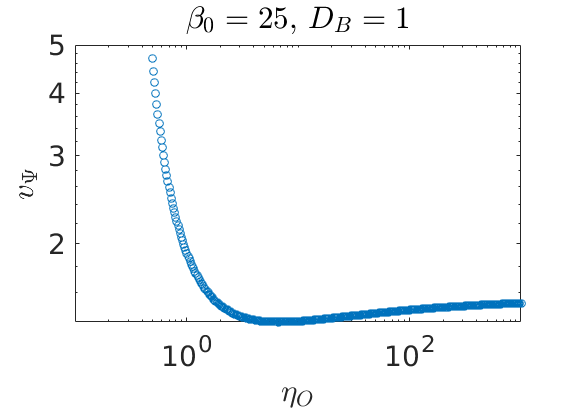}
	\includegraphics[width=0.33\columnwidth]{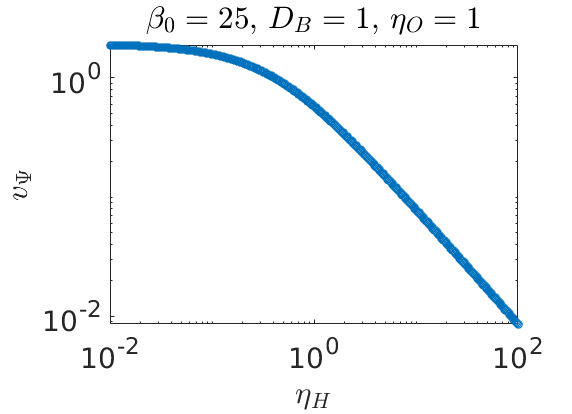}
	\includegraphics[width=0.33\columnwidth]{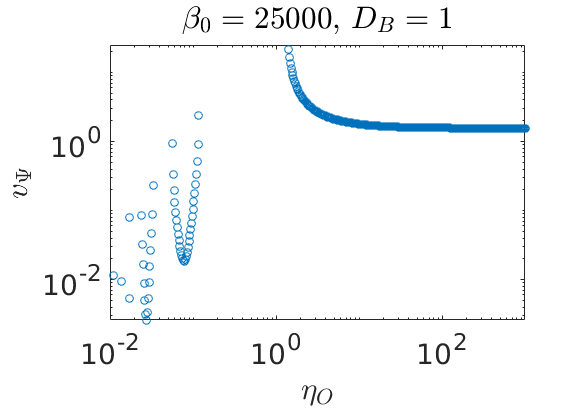}\\
	\includegraphics[width=0.33\columnwidth]{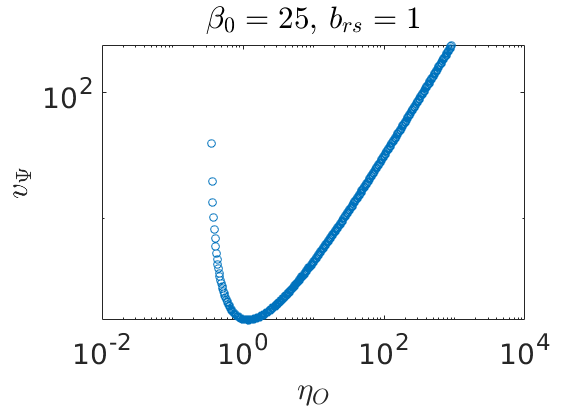}
	\includegraphics[width=0.33\columnwidth]{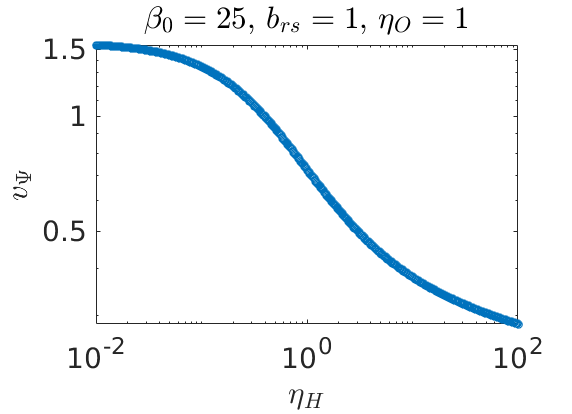}
	\includegraphics[width=0.33\columnwidth]{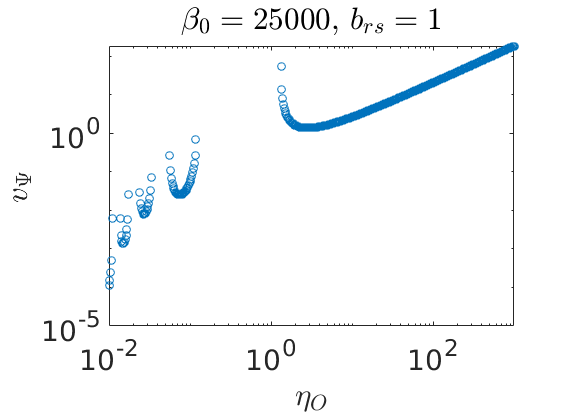}\\
	\includegraphics[width=0.33\columnwidth]{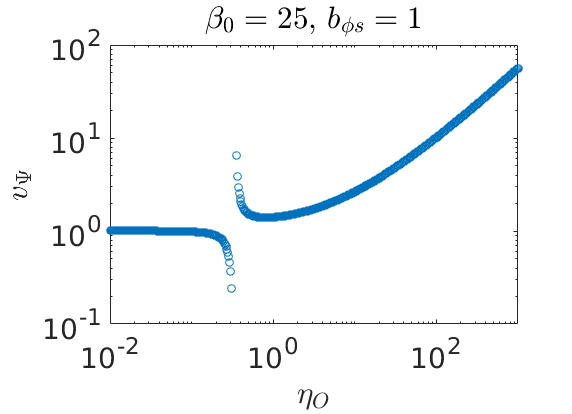}
	\includegraphics[width=0.33\columnwidth]{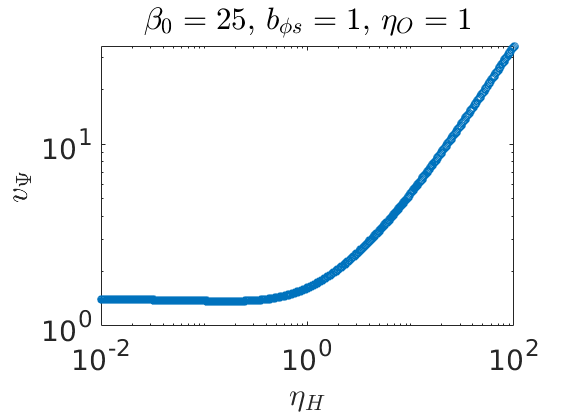}
	\includegraphics[width=0.33\columnwidth]{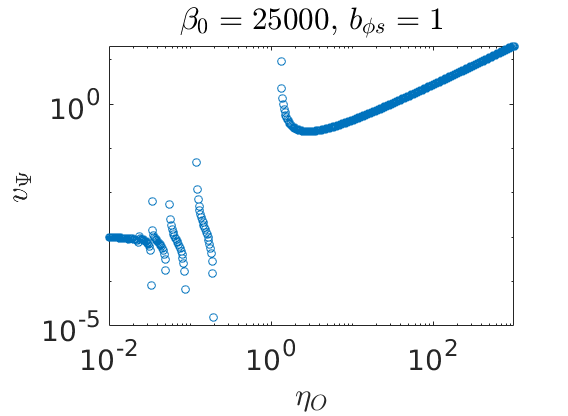}
    \end{center}
    \caption{Plots of $u_\Psi$ (in the plots they are labelled as $v_\Psi$) variation with diffusivities. Each column corresponds to a specific midplane $\beta_0$, with $\beta_0=25,25,2.5\cdot10^4$ from left to right respectively. The first row shows the contribution due to the `Hydro' source term, while the rows below show the contributions due to the $D_B$, $b_{rs}$ and $b_{\phi s}$ source terms as we go down. Values where $u_\Psi<0$ are not plotted due to the logarithmic scales used, but they exist in the unstable regions which are marked by the presence of multiple asymptotes.}
    \label{fig:GO.uPsiplots}
\end{figure*}

Flux transport is again calculated by integrating the radial component of the induction equation:
\begin{equation}\label{eq:GO.uPsi.general}
    u_\Psi = u_r 
    + \Tilde{\eta}_{O}
    (\p_\zeta b_r -D_B) + \Tilde{\eta}_H \p_\zeta b_\phi = \cst,
\end{equation}
where
\begin{equation}
    u_\Psi \equiv \f{r}{H}\f{v_\Psi}{c_s}
\end{equation}
is the dimensionless magnetic flux transport velocity, same as the one defined in equation \eqref{eq:RLVSM.vPsi.def}.

For simplicity, we used constant diffusivity profiles in our calculations. We present here only results from the positive polarity case of $\eta_H (\boldsymbol{B}_z \cdot \boldsymbol{\Omega}) > 0$, as most solutions for the negative polarity case failed to converge, indicative of unstable configurations. Representative plots of $u_\Psi$ variation with diffusivities can be found in Figure \ref{fig:GO.uPsiplots}. For all our solutions, we isolate a particular source term and set its coefficient to $1$. The case for $b_{\phi s} = 1$ corresponds to a torque that spins the disc up, leading to decretion rather than the normal accretion. To examine the case for $b_{\phi s} = -1$, which mimics the effect of a MCW, we simply need to reverse the sign of $u_\Psi$ when interpreting the plots.

\subsubsection{Variation with Ohmic resistivity}

Characteristic plots of $u_\Psi$ variation with $\eta_O$ are displayed in the left and right columns of Figure \ref{fig:GO.uPsiplots}. We found that qualitatively, the `Hydro' and $D_B$ source terms share the same trend, while the $b_{rs}$ and $b_{\phi s}$ source terms follow a different trend. As the midplane field strength is decreased ($\beta_0$ increased), solutions change from a single smooth curve, to multiple curves separated by asymptotes in $\eta_O$. The multiple curves signify the onset of instabilities, and this result is in agreement with the picture given by our model in Section \ref{sec:RLVSM}, with instabilities setting in below a critical field strength. The same is observed for when $\eta_O$ increases. Beyond a threshold diffusivity, the solution is stable and has only one branch.

In the stable configurations, both `Hydro' and $D_B$ source terms contribute to radially outward flux transport ($u_\Psi>0$). Away from the asymptote, which appears to indicate a region of instability, the flux transport velocity increases fractionally ($<10\%$) as $\eta_O$ is increased. 

On the other hand, for both $b_{rs}$ and $b_{\phi s}$ source terms, an increase in $\eta_O$ leads to a similar order of magnitude increase in $u_\Psi$, which is postive in the stable region. This confirms both the picture in our previous model (see Section \ref{sec:RVLSM.EyTrends}) that a positive inclination of the poloidal field away from the star when coupled with diffusivity drives outward flux transport, and also that the rate at which it does so correspond to roughly linear increases. Remembering to reverse the sign for an actual wind, the results for $b_{\phi s}$ show us that a wind, coupled with resistivity, leads to an accretion of flux ($u_\Psi<0$), again in agreement with our results from the shearing box model.

\subsubsection{Variation with Hall diffusivity}

When the Hall term was the only non-ideal MHD effect present, the solver did not converge. Analytically, we can determine that we would have iso-rotation and no flux transport. We therefore examined the cases where a background Ohmic resistivity was also present.

$u_\Psi$ variation with $\eta_H$ are found to share the same trend between the `Hydro', $D_B$ and $b_{rs}$ source terms, while those for the $b_{\phi s}$ source term follow a different trend. Again, as the midplane field strength is decreased, solutions go from a single smooth curve, to multiple curves separated by asymptotes in $\eta_H$. The same is observed for when $\eta_H$ increases. Beyond a threshold diffusivity, the solution is stable and only has one branch.

In the stable configurations, `Hydro', $D_B$ and $b_{rs}$ source terms contribute to a radially outward flux transport ($u_\Psi>0$). Away from the final asymptote that marks the region of instability, the flux transport velocity decreases to zero as $\eta_H$ is increased. This is similar to the picture in our previous shearing box model that indicates a decrease in the inclination driven flux transport rate with an increase in the Hall coefficient, with the only difference being that the nonlinear Hall term in our previous model leads only to a minor correction of the flux transport rate, whereas the linearised Hall term here reduces it to zero. On the other hand, for the $b_{\phi s}$ source term, an increase in $\eta_H$ leads to a similar order of magnitude increase in $u_\Psi$. $u_\Psi$ is positive in the stable region. Therefore for $b_{\phi s} = -1$, mimicking the effect of a MCW, increasing the Hall contribution leads to a significant accretion of flux ($u_\Psi<0$), which is again in agreement with our results from the shearing box model. 

\subsection{Analytic models}

To help us interpret the trends in flux transport, we developed three approximate analytic models following the same procedure as Section 4.1 and Appendix A of GO1. Full mathematical details of how these three models were calculated can be found in Appendix D (see online supplementary materials), but here we give a brief qualitative description of the procedure and assumptions used to derive these models.

We split the disc into two regions: a passive field region with weak (passive) magnetic field ($\beta\gg1$) around the midplane where hydrodynamic effects dominate over magnetic effects;
then further up the disc, we have a region with strong magnetic field ($\beta\ll1$) where magnetic effects are dominant and the field is approximately force-free. The transition point between the regions is where the magnetic pressure is equal to the thermal pressure, and is given by:
\begin{equation}
    \zeta_B = \sqrt{\ln{\left( \f{2}{\pi} \beta_0^2 \right)}}.
\end{equation}
We can see from this that the lower the field strength, the larger the value of $\zeta_B$, and the better the approximation of the midplane region as under a `passive' field.

The models are constructed by first calculating the general forms of the approximate analytic profiles in each of the two regions, constrained by the given midplane and disc surface ($\zeta\to\infty$) boundary conditions. We assume that the Lorentz force is negligible in the passive field region hence ignore the effect of the magnetic field in the velocity profile there. In the force-free region, nothing can compensate the Lorentz force, and the fluid is frozen into the magnetic field lines, with iso-rotation being enforced as in the ideal MHD case. The two regions are then connected across the transition region, which is assumed to be infinitesimally thin about $\zeta_B$. When connecting the regions, we assumed continuity of the magnetic field components across the transition, and integrated the horizontal components of the induction equation over the transition region to find the appropriate boundary conditions. 

The first two of the models represent different diffusivity regimes: the first assumes only the presence of a constant Ohmic diffusivity, while the second addresses the situation when the disc is dominated by Hall drift but with a small Ohmic contribution. Our third model improves on the first model by using better boundary conditions that more appropriately address the presence of the intermediate transition region, and is derived using the description outlined in Appendix A of GO1. 

\subsubsection{Source terms and disc vertical profiles}


\begin{figure*}
    \begin{center}
	\includegraphics[width=0.34\columnwidth]{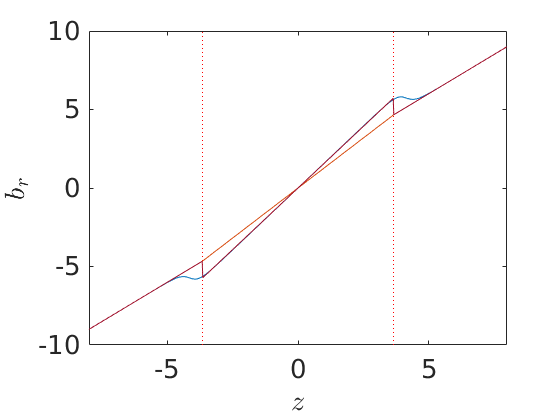}
	\includegraphics[width=0.34\columnwidth]{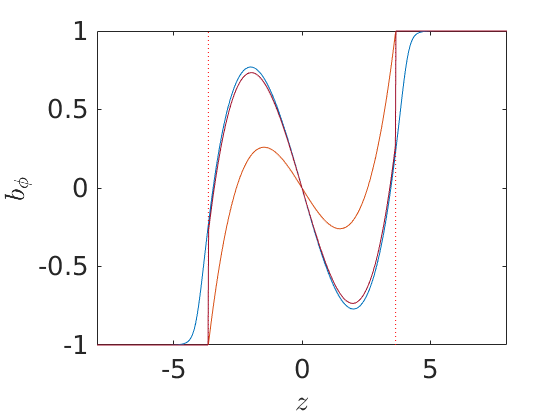}\\
	\includegraphics[width=0.34\columnwidth]{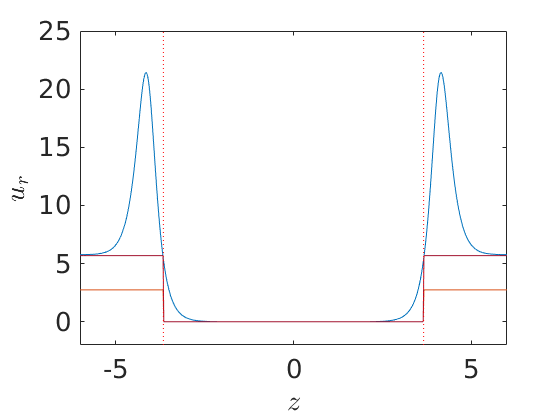}
	\includegraphics[width=0.34\columnwidth]{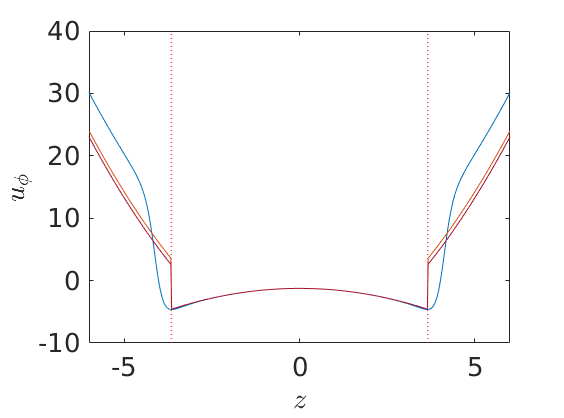}
    \end{center}
    \caption{Vertical structure profiles of the radial (left-hand panels) and azimuthal (right-hand panels) magnetic field (top panels) and velocity (bottom panels) for the Ohm only case with $\eta_O = 10$ and $\beta_0=1000$. We used the 'Hydro' source terms, and also set $b_{rs}$, $b_{\phi s}$ and $D_B$ to 1. Red lines correspond to the simple two-zone analytic model (see Appendix D1 in the online supplementary materials), while purple lines include improved boundary conditions accounting for the transition region (see Appendix D3 in the online supplementary materials). The vertical dotted lines mark the height $\zeta_B$ where the transition between passive and force-free field regions take place.}
    \label{fig:GO.OhmOnly.profiles}
\end{figure*}

\begin{figure*}
    \begin{center}
	\includegraphics[width=0.34\columnwidth]{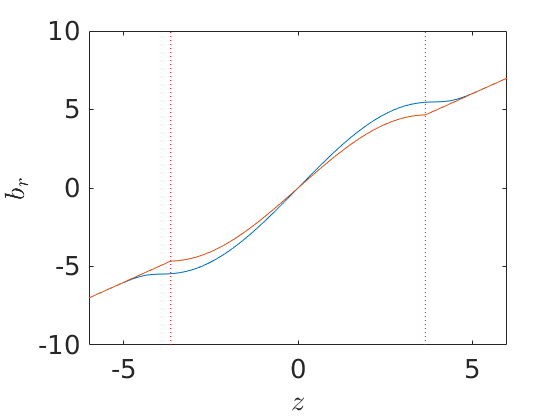}
	\includegraphics[width=0.34\columnwidth]{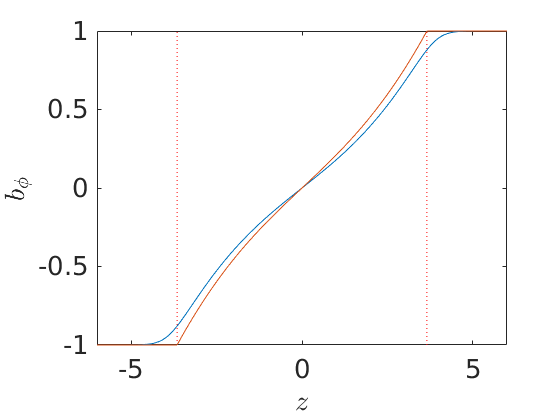}\\
	\includegraphics[width=0.34\columnwidth]{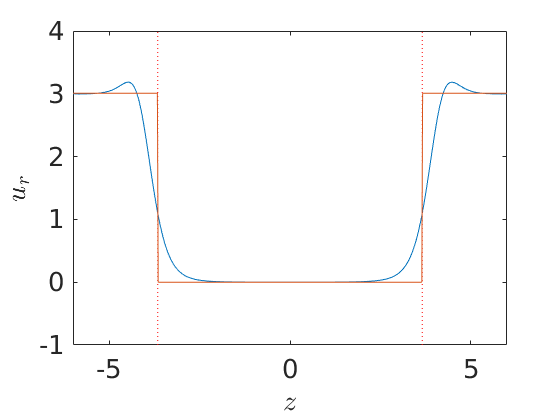}
	\includegraphics[width=0.34\columnwidth]{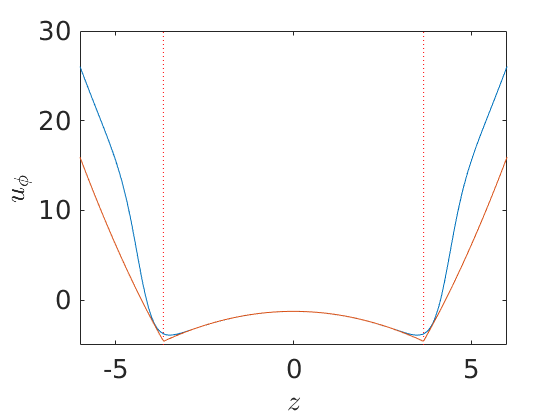}
    \end{center}
    \caption{Same as Figure \ref{fig:GO.OhmOnly.profiles} but for the Hall dominated case with $\eta_O = 1$, $\eta_H = 10$. Red lines correspond to the two-zone analytic model (see Appendix D2 in the online supplementary materials). }
    \label{fig:GO.HallDom.profiles}
\end{figure*}


Figures \ref{fig:GO.OhmOnly.profiles} and \ref{fig:GO.HallDom.profiles} show the vertical structure profiles in the case of a disc with only Ohmic diffusion and a Hall dominated disc with small Ohmic contribution respectively. The red lines are from the simple two-zone models that use the boundary condition of continuous magnetic fields, while the purple lines in the Ohm only case are from the improved two-zone model that take into account the intermediate region, and modifies the jump condition in the magnetic fields at the transition point.
They both provide good qualitative descriptions of the solution, with the improved model matching very well to the actual numerical solution.

By analysing the mathematical forms of the analytic solutions (see Appendix D of the online supplementary materials), we can deduce how the various source terms affect the shape of the disc vertical profile in the different variables. The radial magnetic gradient $D_B$ provides the background gradient in $\zeta$ for the profile of $b_r$, while $b_{rs}$ sets the limit of $b_r(\infty)$ when $D_B$ is absent, defining the surface inclination of the poloidal field. Similarly, $b_{\phi s}$ sets the value of $b_\phi(\infty)$, and hence the magnetic torque acting on the disc. Ohmic resistivity causes bends to occur in the passive field region for the $b_\phi$ profile, while the Hall drift, which operates at $\pi/2$ phase to Ohmic diffusion, causes the bends to happen in $b_r$ instead. $u_r$ tends towards the flux transport value as $\zeta\to\infty$, but is largely zero in the passive field region. $u_\phi$ is similarly very small in the disc, but increases drastically in the force-free region as a result of iso-rotation with the magnetic fields. 

\subsubsection{Flux transport rates}

\begin{figure*}
    \begin{center}
	\includegraphics[width=0.42\columnwidth]{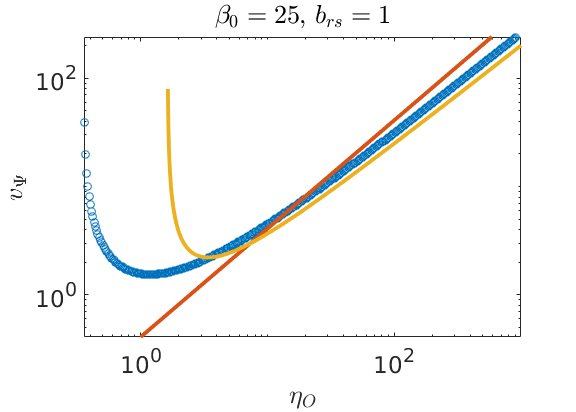}
	\includegraphics[width=0.42\columnwidth]{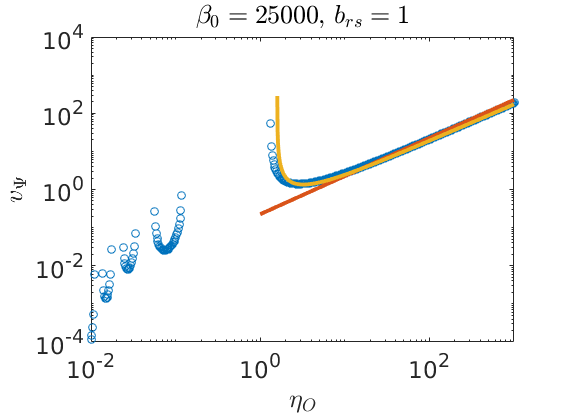}\\
	\includegraphics[width=0.42\columnwidth]{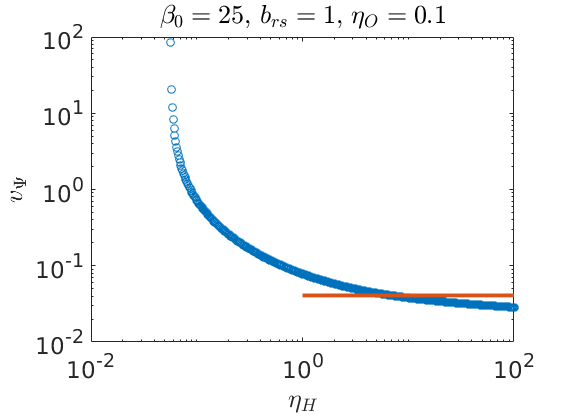}
	\includegraphics[width=0.42\columnwidth]{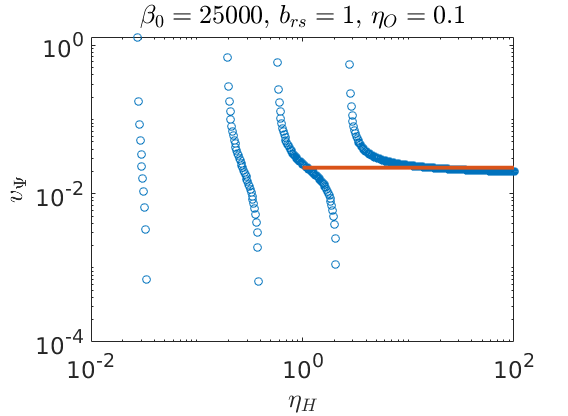}\\
	\includegraphics[width=0.42\columnwidth]{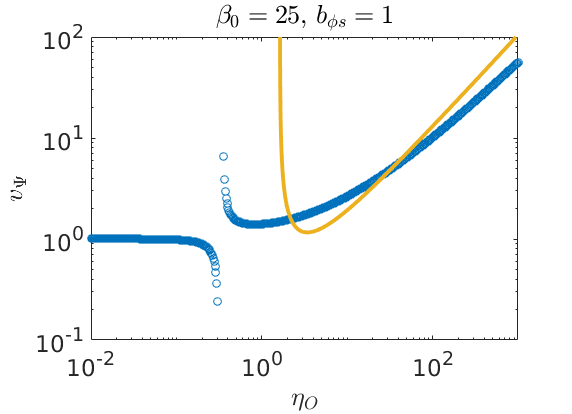}
	\includegraphics[width=0.42\columnwidth]{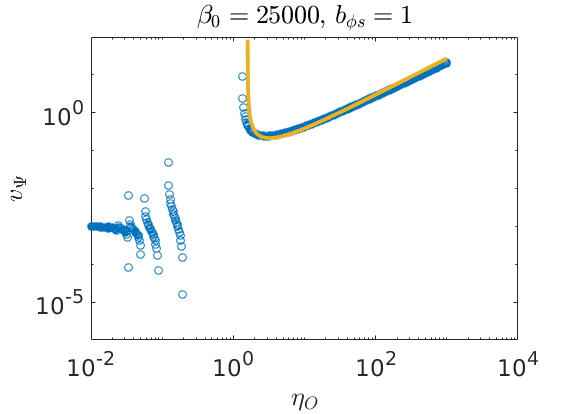}\\
	\includegraphics[width=0.42\columnwidth]{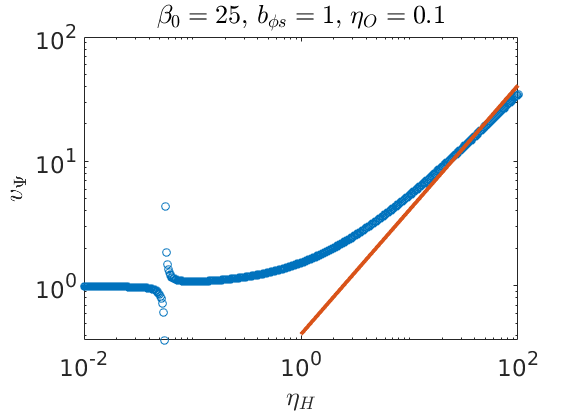}
	\includegraphics[width=0.42\columnwidth]{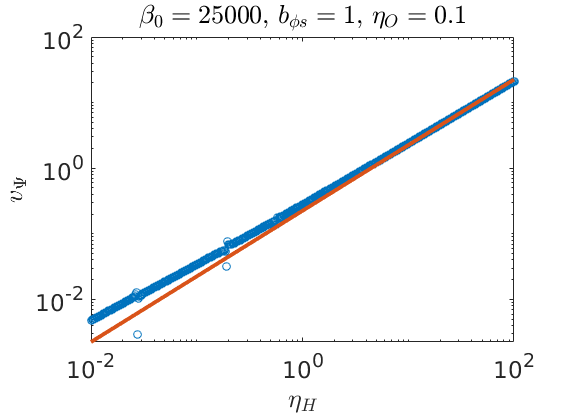}
    \end{center}
    \caption{Plots of radial flux transport rates given by $v_\Psi$ and their variation with Ohmic (first and third rows) and Hall (second and fourth rows) diffusivities, coupled with $b_{rs}=1$ (top two rows) and $b_{\phi s}=1$ (bottom two rows) source terms, at strong ($\beta_0=25$, left hand column) and weak ($\beta_0=25000$, right hand column) magnetisations. Red lines correspond to the two-zone analytic models for Ohm only (first row) and Hall dominated (second and fourth rows) cases, while orange lines include improved boundary conditions accounting for the transition region (first and third rows). }
    \label{fig:GO.uPsi.analytic}
\end{figure*}

Here, we analyse only the flux transport driven by inclination and outflow, while we leave those due to large scale radial gradients to a future investigation. Analytically, we find that the flux transport is given by these simple expressions:
\begin{equation}\label{eq:GO.uPsi.Ohm.analytic.exp}
    u_\Psi = 
    \eta_O \f{b_{rs}}{\zeta_B} 
\end{equation}
for the Ohm only case,
\begin{equation}\label{eq:GO.uPsi.HallDom.analytic.exp}
    u_\Psi = 
    \eta_H \f{b_{\phi s}}{\zeta_B}
    + \eta_O \f{b_{rs}}{\zeta_B}
\end{equation}
for the Hall dominated case, and
\begin{equation}\label{eq:GOHallDom.Ey.analytic}
    u_\Psi 
    = \f{\eta_O}{\zeta_B\left(
    1 - \f{\pi}{2\eta_O} \right)
    - \f{1}{\zeta_B}\ln{\left(\f{2}{\eta_O}\right) } }
    \left\{ b_{rs}
    + \f{\pi b_{\phi s}}{\zeta_B^2}
    \right\}
\end{equation}
for the Ohm only case with improved boundary conditions accounting for the transition region.

These expressions do not involve the radial advection velocity in the ideal MHD limit like second term in Equation \eqref{eq:RLVSM.Ey.analytic} 
because of the separation of the disc into two zones, with the flux transport value determined by the expression in the passive field region where the only non-diffusive contribution, $u_r$, is approximated to be zero. In the absence of any diffusivity, it can be easily shown that the original equations admit an analytic solution with uniform $u_r$ throughout the disc, and the flux transport is simply due to advection via the wind torque, with $u_\Psi = u_r =4 b_{\phi s}/\beta_0$. Hence our simple two-zone analytic models should only be used in the high diffusivity regime, as is common in PPDs.

As in the uniform density shearing box model (see Section \ref{sec:RLVSM.cstdensitymodel}), we find Ohmic diffusivity to be linked with inclination, and Hall diffusivity with outflow, in driving flux transport. Again, it is worth noting that the $\eta_H$ term in Equation \eqref{eq:GO.uPsi.HallDom.analytic.exp} is essentially of the same form as Equation 10 of \citep{BaiStone2017}. In the modified analytic model where jump conditions in the magnetic fields are calculated by taking the transition region into account, we have a modified flux transport rate which also couples Ohmic resistivity with outflow.

Figure \ref{fig:GO.uPsi.analytic} shows how these analytic predictions fit the actual flux transport rates calculated. We first look at the inclination cases which are the top two rows of the figure.
Here the red lines are from the simple two zone models, while the orange lines in the Ohm only cases are from the third model that includes the modified boundary conditions.
We see that the analytic models offer good descriptions in the limit of weak field ($\beta_0$ large) and high diffusivity. This is because under these conditions, the passive field region is extended and better matched by the assumptions used in the analytic model. 
Particularly, the asymptotes of the modified analytic model appears to be able to predict the transition point to instability, which occurs when the denominator of Equation \eqref{eq:GOHallDom.Ey.analytic} is equal to zero. It would be worth investigating in the future how the region of instability is influenced by the balance between diffusive effects (from $\eta_O$) and the field strength (from $\zeta_B$).

Next, we turn to the outflow cases, which are displayed in the bottom two rows of Figure \ref{fig:GO.uPsi.analytic}.
The orange lines in the top two Ohm only plots are from the modified analytic model,
while the red lines in the bottom two Hall dominated panels are from the simple analytic model.
We can see that again they predict the outflow driven flux transport fairly well in the high diffusivity, weak field ($\beta_0$ large) regime. The simple two zone models are unable to describe the presence of the asymptote and transition to the unstable regime, but the model with modified boundary condition matches the first asymptote in the Ohm-only case in the weak field limit. 

The relative success of the analytic models so far (though with limitations, such as in the outflow case) in matching the numerical solutions suggest that it is also possible to understand the flux transport driven by other source terms similarly, and we will be looking at constructing these models and examining their interpretations in the future.

\section{Discussion}

\subsection{Comparing the shearing box and Guilet \& Ogilvie models}

Both the shearing box and the GO models share the same qualitative trends of how diffusivities, coupled with inclination and outflow, can be effective in driving flux transport. In particular, approximate analytic solutions for both models under simplified schemes show how Ohmic resistivity (and the `Ohm-like' term of ambipolar diffusivity) is coupled with inclination to drive radially outward flux transport, while Hall drift with $\eta_H (\boldsymbol{B}_z \cdot \boldsymbol{\Omega}) > 0$ is coupled with outflow in facilitating radially inward flux transport. The similarity between the expressions in Equations \eqref{eq:RLVSM.Ey.analytic}, \eqref{eq:GO.uPsi.Ohm.analytic.exp} and \eqref{eq:GO.uPsi.HallDom.analytic.exp} suggest that we can identify $\zeta_B$, the height at which magnetic pressure equals thermal pressure, as the relevant value for the disc height $H$ to be used in the constant density shearing box model of Section \ref{sec:RLVSM.cstdensitymodel}. Both models also show similar trends and behaviour in how the stability of solutions vary with diffusivity and magnetisation values. The large similarities between the results of the two models may be due to the fact that horizontal magnetic field components are largely relatively small compared with the vertical field for the parameter space explored, hence the effects due to the nonlinearity of the Hall and ambipolar terms are less significant overall. 

\subsection{Comparison with current global simulations}

While outward inclination coupled with Ohmic resistivity has long been known to facilitate outward flux transport, the possibility of Hall drift coupled with an outflow driving significant flux transport has only been briefly noted \citep{Bai_2014} in the past, and never extensively investigated. Only recently have global simulations \citep{Bethuneetal2017,BaiStone2017,Suriano_etal_2017,Suriano_etal_2018} begun exploring the flux transport problem in laminar PPDs where all three non-ideal effects are accounted for. In \citet{BaiStone2017}, the mechanism governing flux evolution can be described as a competition between the Hall effect, ambipolar diffusion, and the MCW-driven accretion. In the positive (negative) polarity configuration, the Hall effect transports flux (out)inwards rapidly in the midplane regions and (in)outwards slowly in the disc upper layers. Ambipolar diffusion, on the other hand, always transports flux outward, whereas MCW driven accretion advects flux inwards. In the aligned case, there is a cancellation effect between the inward flux transport due to Hall and the MCW and outward due to ambipolar, while all three effects work to transport flux outward in the negative polarity case. Our calculations also found a similar picture in that Hall drift reduces the effect of outward flux transport due to ambipolar/Ohmic diffusion in the positive polarity case. However, it is unclear from the results of \citet{BaiStone2017} whether there is any specific coupling between Hall drift and the wind that further enhances flux transport inwards. Our models also agree with \citet{Bethuneetal2017} and \cite{BaiStone2017} that the polarity of the magnetic field is a significant parameter on disc dynamics and flux transport when Hall drift is present. However, we are not able to confirm the flux transport trends in the negative polarity cases due to the breakdown of stable solutions in our models. Our model restricts the overall flux transport rate in a radially local region to be constant across all scale-heights, which is indeed what is found in global simulations \citep{BaiStone2017,Bai_2017}. 

A significant proportion of the parameter space explored yielded unstable solutions characteristic of MRI channel modes, and it would be interesting to see in future studies if they might relate to flux transport mechanisms reported in global simulations that are cyclic in nature, such as those in \citet{Suriano_etal_2017,Suriano_etal_2018,Suriano_etal_2019}.

\subsection{Implications, model limitations and future directions}

One important implication of our results is the need to further investigate the interplay between Hall drift and a wind outflow in facilitating flux transport. \citet{Bai_2016}, which attempted to present a framework of global PPD evolution incorporating the latest advances in PPD research, found disc evolution to be largely dominated by wind-driven processes, while viscous spreading is suppressed. He noted that the timescale of disc evolution is largely governed by the global flux evolution in the disc, and admitted that an understanding of this and in particular of how the Hall effect affects magnetic flux evolution is still lacking. To bridge this gap, a more detailed way of modelling the wind outflow in the presence of the Hall effect than our simple presciption of a non-zero surface $B_\phi$ would be required, where disc boundary conditions are formally matched to the Blandford \& Payne solutions, and where we allow for a vertical outflow $v_z$. The feasibility of such approach, tailored for the flux transport problem, will be examined in future work. 

The question of how to model asymmetrical solutions semi-analytically should also be looked into. Disc solutions with vertical symmetry breaking have been observed in many recent global disc simulations \citep{Bethuneetal2016,Bethuneetal2017,Bai_2017}. They have drastic contrasts in their overall accretion and flux transport rates compared to their standard symmetric counterparts. It would be worth investigating the effect of symmetry breaking or having the opposite symmetry to the one prescribed in the current work using our semi-analytic approach in future work. Currently, switching the even-odd symmetry of the variables would yield the trivial result of zero net flux transport in our local models, as the transport from the bottom and top parts of the disc will cancel each other out. Modelling the flux transport meaningfully in such context therefore would require careful thought.

Future investigations should also seek to implement more realistic diffusivity profiles, such as those used in these simulations \citep{Bethuneetal2016,Bethuneetal2017,Bai_2017}. The flux transport mechanism reported in \citet{Suriano_etal_2018} is heavily dependent on the variation on the ambipolar diffusivity in the disc, and involves a time-dependent periodic flux concentration processes. 

Given the theoretical work nature of the present work, shearing box simulations would be needed to verify the equilibrium solutions found by our semi-analytic models. The advantage of running a simulation is that it gives us insight into how the disc develops over time before reaching equilibrium, and it would be interesting to see whether any structures with periodic cycles are formed in the unstable regimes, and how they may inform us about the flux transport mechanisms reported in global simulations.

Finally, it would be good to combine these local models into a global flux evolution framework, such as done by \citet{GuiletOgilvie2014, Okuzumi_etal_2014} and \citet{TakeuchiOkuzumi_2014}. The main challenge resides in the fact that Hall/wind-driven flux transport requires knowledge of the disc surface toroidal field, while the previous models only required the calculation of the poloidal field structure which is more readily determined through the Biot-Savart law. Careful modelling of the global wind solution is required, and will be looked into in a future investigation. In the absence of a magnetic wind, the Hall effect (when $\eta_H (\boldsymbol{B}_z \cdot \boldsymbol{\Omega}) > 0$) reduces the radially outward flux transport rate induced by Ohmic and/or ambipolar diffusion coupled with an inclined field. Hence we should expect the PPDs where the Hall effect is present, field is aligned with rotation, and where no outflow is launched, to settle into a more highly magnetised steady-state solution than those previously calculated.

\section{Summary and conclusions}

In this work, we have developed a formalism to compute the radially local effects of non-ideal diffusivities coupled with inclination, outflow and large scale radial gradients on disc dynamics through the use of a semi-analytic multiscale asymptotic approach. We investigated the flux transport due to inclination and outflow first at the shearing box order, and then at the Guilet \& Ogilvie order \citep{GuiletOgilvie2012}, which also allows the computation of the additional contributions from large scale radial gradients. Our findings from both models are qualitatively similar, and we examined the trends of inclination and outflow driven flux transport and their variation with the three diffusivities. Using approximate analytic models, we gained insights into how diffusivities are coupled with the various other parameters in facilitating flux transport, and derived simple relations for estimating this.

Our main findings include:
\begin{enumerate}
    \item Stable disc configurations arise from having a strong field (low $\beta_0$) and high diffusivity values, while weak field and low diffusivities give rise to unstable configurations characteristic of MRI channel modes. 
    \item In the positive polarity case, where $\eta_H (\boldsymbol{B}_z \cdot \boldsymbol{\Omega}) > 0$, all diffusivities are stabilising, while the Hall effect becomes destabilising in the anti-aligned case.
    \item Outward inclination of the poloidal field coupled with Ohmic and ambipolar diffusion both drive radially outward flux transport which increases roughly linearly with diffusivity, while the Hall effect coupled with outward inclination reduces the flux transport rate but does not reverse its direction.
    \item When the toroidal field in the disc is small compared to the overall field, outflow driven flux transport is not significantly affected by the presence of Ohmic and ambipolar diffusion, but is significantly enhanced when the Hall effect is present. In the high $\beta_0$, high diffusitivity limit common to PPDs, this scales roughly linearly with $\eta_H$ and the surface toroidal field strength. All outflow driven flux transport is radially inward, as would be expected from advection due to the accretion flow caused by vertical removal of angular momentum in the outflow. When the toroidal field in the disc is large, the flux transport behaviour with Ohmic and Hall coefficients remain the same, while it is reversed in the ambipolar case, as the "Hall-like" component of the ambipolar term dominates over the "Ohm-like" component, and acts in the opposite direction to the Hall term.
\end{enumerate}

This work represents an initial effort toward modelling flux transport in PPDs incorporating all three non-ideal effects, inclination of the large scale field, outflow and the presence of large-scale radial gradients. At present, we have focused on the local transport in the disc, with a range of parameters characterising the contribution from each effect, and assumed a quasi-steady equilibrium state. Future work would need to address the time-dependent aspect of disc evolution, and also consider the flux transport globally under more realistic physical parameters for better comparison and understanding to the results of present and future PPD simulations.

\section*{Acknowledgements}

We would like to thank the annonymous referee for a detailed report which improved the presentation of the manuscript. PKCL would like to thank J\'er\^ome Guilet for useful discussions, and for letting him have a copy of the code used for the 2012 Guilet \& Ogilvie paper. He would also like to thank the Croucher Foundation and the Cambridge Commonwealth, European \& International Trust for their generous support in funding his PhD studentship through a Cambridge Croucher International Scholarship.




\bibliographystyle{mnras}

\appendix

\section{Multiscale approach equations at different orders}
\label{app:MultiscaleOrderEqns}

\subsection{Mass and momentum equations}

The mass and momentum equations, assuming axisymmetry, are:
\begin{equation}
    \rmD \rho = -\rho \Delta,
\end{equation}
\begin{equation}
  \rmD u_r-\f{u_\phi^2}{r}=-\p_r\Phi-\f{1}{\rho}\p_r p + \f{1}{\mu_0} \left( J_\phi B_z - J_z B_\phi \right),
\end{equation}
\begin{equation}
  \rmD u_\phi+\f{u_r u_\phi}{r}= \f{1}{\mu_0} \left( J_z B_r - J_r B_z \right),
\end{equation}
\begin{equation}
  \rmD u_z=-\p_z\Phi-\f{1}{\rho}\p_z p + \f{1}{\mu_0} \left( J_r B_\phi - J_\phi B_r \right),
\end{equation}
where
\begin{equation}
  \rmD=\p_t+u_r\p_r+u_z\p_z,
\end{equation}
\begin{equation}
  \Delta=\f{1}{r}\p_r(r u_r)+\p_z u_z,
\end{equation}
and the current is given by
\begin{equation}
    \bmJ = \nabla\times \bmB.
\end{equation}

We introduce the following operators:
\begin{equation}
  \rmD=\rmD_0+\epsilon\rmD_1+\epsilon^2\rmD_2+\cdots,
\end{equation}
with
\begin{equation}
  \rmD_0=\p_t+v_{r0}\p_\xi+v_{z0}\p_\zeta,
\end{equation}
\begin{equation}
  \rmD_1=v_{r1}\p_\xi+v_{r0}\p_r+v_{z1}\p_\zeta,
\end{equation}
\begin{equation}
  \rmD_2=\p_\tau+v_{r2}\p_\xi+v_{r1}\p_r+v_{z2}\p_\zeta.
\end{equation}
Also
\begin{equation}
  \Delta=\Delta_0+\epsilon\Delta_1+\cdots,
\end{equation}
with
\begin{equation}
  \Delta_0=\p_\xi v_{r0}+\p_\zeta v_{z0},
\end{equation}
\begin{equation}
  \Delta_1=\p_\xi v_{r1}+\f{1}{r}\p_r(rv_{r0})+\p_\zeta v_{z1}.
\end{equation}

Assuming the potential of Equation (7), the leading-order equations are then
\begin{equation}
  \rmD_0\rho_0=-\rho_0\Delta_0,
\end{equation}
\begin{equation}
\begin{aligned}
  & \rmD_0v_{r0}-2\Omega v_{\phi0}= -\f{1}{\rho_0}\p_\xi p_0 \\
  &  \qquad \qquad
  + \f{1}{\mu_0} \bigg[ \left( \p_{\zeta} B_{r0}
  - \p_\xi B_{z0} \right) B_{z0} 
  - B_{\phi0} \p_\xi B_{\phi0}  \bigg],
\end{aligned}
\end{equation}
\begin{equation}
\begin{aligned}
  & \rmD_0v_{\phi0}+\f{1}{r}\p_r(r^2\Omega)v_{r0} = 0\\
  & \qquad\qquad
  + \f{1}{\mu_0} \bigg[ \p_\xi B_{\phi0} B_{r0} 
  + B_{z0} \p_\zeta B_{\phi 0} \bigg],
\end{aligned}
\end{equation}
\begin{equation}
\begin{aligned}
  \rmD_0v_{z0}=& -\Psi\zeta-\f{1}{\rho_0}\p_\zeta p_0 \\
  &  + \f{1}{\mu_0} \bigg[ \left( 
  - \p_\zeta B_{\phi 0} \right) B_{\phi0} 
  - \left( \p_{\zeta} B_{r0}
  - \p_\xi B_{z0}  \right) B_{r0} \bigg] ,
\end{aligned}
\end{equation}

At the next order, we have 
\begin{equation}
\begin{aligned}
  \rmD_1\rho_0+\rmD_0\rho_1=-\rho_1\Delta_0-\rho_0\Delta_1,
\end{aligned}
\end{equation}
\begin{equation}\label{r_mom_2ndO}
\begin{aligned}
  & \rmD_1v_{r0}+\rmD_0v_{r1}-\f{v_{\phi0}^2}{r}-2\Omega v_{\phi1} \\
  & \qquad 
  =  -(\p_r\Psi)\f{1}{2}\zeta^2+\f{\rho_1}{\rho_0^2}\p_\xi p_0 
  -\f{1}{\rho_0}(\p_r p_0+\p_\xi p_1)\\
  & \qquad \qquad
  + \f{1}{\mu_0} \left( J_{\phi1} B_{z0} - J_{z1} B_{\phi0}
  + J_{\phi0} B_{z1} - J_{z0} B_{\phi1} \right),
\end{aligned}
\end{equation}
\begin{equation}
\begin{aligned}
  & \rmD_1v_{\phi0}+\rmD_0v_{\phi1}+\f{v_{r0}v_{\phi0}}{r}+\f{1}{r}\p_r(r^2\Omega)v_{r1}\\
  & \qquad 
  = \f{1}{\mu_0} \left( J_{z1} B_{r0} - J_{r1} B_{z0} 
  + J_{z0} B_{r1} - J_{r0} B_{z1} \right),
\end{aligned}
\end{equation}
\begin{equation}
\begin{aligned}
  & \rmD_1v_{z0}+\rmD_0v_{z1} = 
  \f{\rho_1}{\rho_0^2}\p_\zeta p_0-\f{1}{\rho_0}\p_\zeta p_1 \\
  & \qquad\qquad
  + \f{1}{\mu_0} \left( J_{r1} B_{\phi0} - J_{\phi1} B_{r0} 
  + J_{r0} B_{\phi1} - J_{\phi0} B_{r1} \right),
\end{aligned}
\end{equation}
with
\begin{eqnarray}
  && J_{r0} = 
  - \p_\zeta B_{\phi 0} , \\
  && J_{r1} = 
  - \p_\zeta B_{\phi 1} , \\
  && J_{\phi0} = \p_{\zeta} B_{r0}
  - \p_\xi B_{z0} , \\
  && J_{\phi1} =
  \p_{\zeta} B_{r1}
  - \p_\xi B_{z1}
  - \p_{r} B_{z0} , \\
  && J_{z0} = \p_\xi B_{\phi0} , \\
  && J_{z1} = \p_\xi B_{\phi1} 
  + \f{1}{r} \p_r (r B_{\phi0} ) .
\end{eqnarray}

\subsection{Induction equation}

We assume axisymmetry throughout this analysis. The full induction equation is given by
\begin{equation}\label{inductionFull}
    \begin{aligned}
        \frac{\partial \boldsymbol{B}}{\partial t}
        + (\boldsymbol{u} \cdot \nabla) \boldsymbol{B}
         = &
        (\boldsymbol{B} \cdot \nabla) \boldsymbol{u}
        - \boldsymbol{B} (\nabla \cdot \boldsymbol{u})
        -\nabla\times\left[\eta_O \bmJ\right] \\
        & - \nabla \times \left[ \eta_H \bmJ \times \bmb \right] 
        + \nabla \times \left\{ \eta_A \left[ \bmJ \times\bmb\right] \times \bmb \right\},
    \end{aligned}
\end{equation}
where $\bmb = \bmB / \lvert \bmB \rvert$, and $\eta_O,\eta_H,\eta_A$ are the Ohmic, Hall and ambipolar diffusivities respectively.

We define
\begin{equation}
    \rmM_B=\rmM_{B0}+\epsilon\rmM_{B1}+\epsilon\rmM_{B2}+\cdots,
\end{equation}
with
\begin{equation}
    \rmM_{B0}=B_{r0}\p_\xi+B_{z0}\p_\zeta,
\end{equation}
\begin{equation}
  \rmM_{B1}=B_{r1}\p_\xi+B_{r0}\p_r+B_{z1}\p_\zeta,
\end{equation}
\begin{equation}
  \rmM_{B2}=B_{r2}\p_\xi+B_{r1}\p_r+B_{z2}\p_\zeta.
\end{equation}
Note that this is of the same form as $\rmD$ with the exception of $\p_t, \p_\tau$ terms. 

At zeroth order, all the terms cancel out.

The order $\epsilon$ equations are:
\begin{equation}
\begin{aligned}
    \rmD_0B_{r0}=\rmM_{B0}v_{r0}-B_{r0}\Delta_0
\end{aligned}
\end{equation}
\begin{equation}
    \rmD_0B_{\phi0}+\Omega B_{r0}=\rmM_{B0}v_{\phi0}+B_{r0}\p_r(r\Omega)-B_{\phi0}\Delta_0
\end{equation}
\begin{equation}
    \rmD_0B_{z0}=\rmM_{B0}v_{z0}-B_{z0}\Delta_0
\end{equation}

The next order $\epsilon^2$ equations are:
\begin{equation}
\begin{aligned}
    \rmD_0B_{r1}+&\rmD_1B_{r0}-\f{B_{\phi0}v_{\phi0}}{r}-\Omega B_{\phi1}\\
    &=\rmM_{B0}v_{r1}+\rmM_{B1}v_{r0}-\f{v_{\phi0}B_{\phi0}}{r}-B_{r1}\Delta_0-B_{r0}\Delta_1
\end{aligned}
\end{equation}
\begin{equation}
\begin{aligned}
    \rmD_0B_{\phi1}+&\rmD_1B_{\phi0}+\f{B_{r0}v_{\phi0}}{r}+\Omega B_{r1}\\
    &=\rmM_{B0}v_{\phi1}+\rmM_{B1}v_{\phi0}+\f{v_{r0}B_{\phi0}}{r}+B_{r1}\p_r(r\Omega)\\
    & \qquad -B_{\phi1}\Delta_0-B_{\phi0}\Delta_1
\end{aligned}
\end{equation}
\begin{equation}
    \rmD_0B_{z1}+\rmD_1B_{z0}=\rmM_{B0}v_{z1}+\rmM_{B1}v_{z0}-B_{z1}\Delta_0-B_{z0}\Delta_1
\end{equation}

We also have the zero divergence condition
\begin{equation}
    \div{\bmB} = 0,
\end{equation}
which when written in cylindrical coordinates, give:
\begin{equation}
    \p_z B_z=-\p_r(r B_r).
\end{equation}
The resulting equations in increasing order are therefore:
\begin{equation}
    \p_\zeta B_{z0}=-\p_\xi B_{r0},
\end{equation}
\begin{equation}
    \p_\zeta B_{z1}=-\p_\xi B_{r1}-\f{1}{r}\p_r(rB_{r0}).
\end{equation}

\section{Equations for numerical solver in the 1D radially local vertical structure model}
\label{app:RLVSM.num.solve.eqns}

For the numerical solver, we recast Equations (26) to (30) to
\begin{equation}
    \rho' = c_s^{-2}[- \rho \Omega^2 z + J_x B_y - J_y B_x],
\end{equation}
\begin{equation}
    B_x' = \mu_0 J_y,
\end{equation}
\begin{equation}
    B_y' = -\mu_0 J_x,
\end{equation}
\begin{equation}
    E_x' = - \f{3}{2} \Omega B_x.
\end{equation}

We define the diffusion constants as the following:
\begin{equation}
    C_H = \f{\eta_H}{\lvert \bmB \rvert} ,
\end{equation}
\begin{equation}
    C_A = \f{\eta_A} {\lvert \bmB \rvert ^2} ,
\end{equation}
such that
\begin{equation}
    \bmE + \bmv\times\bmB = C_O\bmJ
    + C_H\bmJ\times\bmB - C_A(\bmJ\times\bmB)\times\bmB.
\end{equation}
Then the currents are given by
\begin{equation}
\begin{aligned}
     \begin{pmatrix}[2]
           J_x\\
           J_y
     \end{pmatrix} 
     = 
     \f{1}{\rho^2 \det (\mathbf{R})} 
     \begin{pmatrix}[2]
           A_{11} &
          A_{12}\\
        A_{21}&
        A_{22}
     \end{pmatrix}
     \begin{pmatrix}[2]
           E_x \\
           E_y      
     \end{pmatrix},    
\end{aligned}
\end{equation}
with
\begin{equation}
\begin{aligned}
    & \rho^2 \det (\mathbf{R}) = [C_O + C_A (B_z^2 + B_y^2)]
            [C_O + C_A (B_z^2 + B_x^2)]\rho^2\\
            & \qquad
            + [(C_H B_z - C_A B_x B_y)\rho + (2\Omega)^{-1} B_z^2]\\
            & \qquad\qquad
            [ (C_H B_z + C_A B_x B_y)\rho + 2 \Omega^{-1} B_z^2],
\end{aligned}
\end{equation}
\begin{equation}
A_{11} = [C_O + C_A (B_z^2 + B_x^2)]\rho^2 ,
\end{equation}
\begin{equation}
A_{12} = (- C_H B_z + C_A B_x B_y)\rho^2 - (2\Omega)^{-1} \rho B_z^2,
\end{equation}
\begin{equation}
A_{21} = (C_H B_z + C_A B_x B_y)\rho^2 + 2 \Omega^{-1} \rho B_z^2 ,
\end{equation}
\begin{equation}
A_{22} = [C_O + C_A (B_z^2 + B_y^2)]\rho^2 .
\end{equation}

\section{Guilet \& Ogilvie approach details}

\subsection{Non dimensionalisation}
\label{App:GOnondimen}

We used the following definitions:
\begin{equation}
    \Tilde{\rho} \equiv \f{\rho_0 H}{\Sigma} ,
\end{equation}
\begin{equation}
    u_r \equiv \f{r}{H} \f{ v_{r1} }{c_s} ,
\end{equation}
\begin{equation}
    u_\phi \equiv \f{r}{H} \f{ v_{\phi1} }{c_s} , 
\end{equation}
\begin{equation}
    b_r \equiv \f{r}{H} \f{ B_{r1} } {B_{z0} },
\end{equation}
\begin{equation}
    b_\phi \equiv \f{r}{H} \f{ B_{\phi1} }{B_{z0} } .
\end{equation}
A dimensionless vertical spatial coordinate is also defined as
\begin{equation}
    \zeta \equiv z/H .
\end{equation}
This $\zeta$ is different from the rescaled dimensional variable $\zeta$ defined earlier.

We can then define dimensionless laminar diffusivities:
\begin{equation}
    \Tilde{\eta}_{O} = \f{\eta_{O,l}}{c_s H},
\end{equation}
\begin{equation}
    \Tilde{\eta}_{H} = \f{\eta_H}{c_s H},
\end{equation}
\begin{equation}
    \Tilde{\eta}_A = \f{\eta_A}{c_s H}.
\end{equation}


\section{Mathematical derivation of analytic models for analysing GO model}

\label{app:GO.Analytic}

\subsection{Analytic model Ohmic diffusion only}
\label{app:GO.Analytic.OhmOnly}

This section follows the same procedures as Section 4.1 of Guilet \& Ogilvie (2012). We assume constant Ohmic diffusivity throughout the disc.

\subsubsection{Passive field regime}
$\beta \to \infty $ and the equations become:
\begin{equation}\label{AN.hydro.Ohm:1}
    - 2 u_\phi
    =
    \f{3}{2} + D_H - D_{\nu\Sigma}
    + \left( \f{3}{2} -  D_H \right) \zeta^2 ,
\end{equation}
\begin{equation}\label{AN.hydro.Ohm:2}
    u_r 
    = 0 ,
\end{equation}
\begin{equation}\label{An.hydro.Ohm:3}
    \eta_O \p_\zeta^2 b_r 
    = 0 ,
\end{equation}
\begin{equation}\label{An.hydro.Ohm:4}
    - \eta_O \p_\zeta^2 b_\phi 
    - \p_\zeta u_\phi + \f{3}{2} b_r 
    = 0.
\end{equation}
The azimuthal velocity can then be written as:
\begin{equation}\label{AN.Ohm.hydro.uphi}
    u_\phi = u_{\phi0} + u_{\phi2} \zeta^2,
\end{equation}
\begin{equation}
    u_{\phi0} 
    = -\f{1}{2} \left( 
    \f{3}{2} + D_H - D_{\nu\Sigma} \right),
\end{equation}
\begin{equation}
    u_{\phi2}
    = \f{1}{2}\left( D_H - \f{3}{2} \right).
\end{equation}
From \eqref{An.hydro.Ohm:3}, and the boundary condition $b_r(0)=0$, we can deduce that:
\begin{equation}\label{AN.Ohm.hydro.br}
    b_r = b_{r1} \zeta.
\end{equation}
We can then find $b_\phi$ by substituting \eqref{AN.Ohm.hydro.uphi} and \eqref{AN.Ohm.hydro.br} into \eqref{An.hydro.Ohm:4}, giving us:
\begin{equation}
    \p_\zeta^2 b_\phi 
    = \f{1}{\eta_O}\left( - 2 u_{\phi2} + \f{3}{2} b_{r1} \right) \zeta.
\end{equation}
This, with the boundary condition $b_\phi(0)=0$, has the solution
\begin{equation}\label{AN.Ohm.hydro.bphi}
    b_\phi 
    = \f{1}{6\eta_O}\left( - 2 u_{\phi2} + \f{3}{2} b_{r1} \right) \zeta^3 + b_{\phi1}\zeta.
\end{equation}
The flux transport in this region is given by Equation (82):
\begin{equation}\label{AN.Ohm.hydro.uPsi}
    u_\Psi = \eta_O (\p_\zeta b_r - D_B).
\end{equation}

\subsubsection{Force-free regime}
\label{app:GO.Analytic.OhmOnly.forcefree}

$\beta\ll1$, nothing can compensate the Lorentz force resulting in force-free magnetic fields with current parallel to the field lines. The equations become:
\begin{equation}\label{AN.cor.Ohm:1}
    \p_\zeta b_r
    =
    D_B ,
\end{equation}
\begin{equation}\label{AN.cor.Ohm:2}
    \p_\zeta b_\phi
    = 0 ,
\end{equation}
\begin{equation}\label{AN.cor.Ohm:3}
    \p_\zeta u_r 
    = 0 ,
\end{equation}
\begin{equation}\label{AN.cor.Ohm:4}
    - \p_\zeta u_\phi + \f{3}{2} b_r 
    = 0.
\end{equation}

For our case where the field is almost vertical, the radial and azimuthal currents vanish. Diffusive effects no longer have any importance, and:
\begin{equation}\label{AN.Ohm.forcefree.2}
    \p_\zeta b_\phi = 0,
\end{equation}
\begin{equation}\label{AN.Ohm.forcefree.1}
    \p_\zeta b_r - D_B = 0.
\end{equation}
With our boundary conditions at infinity, we find that throughout the force-free region:
\begin{equation}\label{AN.Ohm.forcefree.bphi}
    b_\phi = \pm b_{\phi s},
\end{equation}
\begin{equation}\label{AN.Ohm.forcefree.br}
    b_r = \pm b_{rs} + D_B \zeta,
\end{equation}
where $\pm$ stands for the sign of $\zeta$.

For the azimuthal velocity, substituting \eqref{AN.Ohm.forcefree.br} and \eqref{AN.Ohm.forcefree.bphi} to \eqref{AN.cor.Ohm:4}:
\begin{equation}
    u_\phi 
    = \f{3}{2}\left(
    b_{rs} \lvert \zeta \rvert 
    + \f{D_B}{2} \zeta^2 \right)
    + u'_{\phi 0},
\end{equation}
where $u'_{\phi 0}$ is an integration constant to be determined.

From Equation (82), the radial velocity is equal to the flux transport velocity and is a constant:
\begin{equation}
    u_r = u_\Psi.
\end{equation}

\subsubsection{Two-zone model}
\label{app:GO.Analytic.OhmOnly.twozone}

The height at which the magnetic pressure is equal to the thermal pressure is given by:
\begin{equation}
    \zeta_B = \sqrt{\ln{\left( \f{2}{\pi} \beta_0^2 \right)}}.
\end{equation}
We assume that for $\zeta<\zeta_B$ we have a `passive field', and for $\zeta>\zeta_B$, we have a `force-free field'. 
We neglect the thickness of the transition where $\beta$ is of order unity, and determine proper boundary conditions at $\zeta=\zeta_B$.
Four conditions are needed to constrain the four unknowns.

Two boundary conditions can be obtained from the analysis of the induction equation. Using \eqref{AN.Ohm.hydro.uPsi}, \eqref{AN.Ohm.hydro.br} and \eqref{AN.Ohm.hydro.bphi} at $\zeta=0$, we obtain:
\begin{equation}
    u_\Psi = \eta_O \left[\p_\zeta \left(b_{r1}\zeta\right) - D_B\right]
    = \eta_O \left(b_{r1} - D_B\right) .
\end{equation}

The azimuthal component of the induction is more complicated because $b_r$ acts as a source term. We integrate between two heights $\zeta_1$ and $\zeta_2$ to get:
\begin{equation}
    \left[ \eta_O \p_\zeta b_\phi 
    + u_\phi 
    \right]^{\zeta_2}_{\zeta_1}
    = 
    \f{3}{2} \int^{\zeta_2}_{\zeta_1} b_r \rmd \zeta.
\end{equation}

We can use this relation between $\zeta_B^{-}$ and $\zeta_B^{+}$ to connect the two regions. We neglect the width of the intermediate region, so the RHS of the equation vanishes. This is equivalent to the condition that horizontal electric field components are continuous across the boundary. Including the force-free conditions that $\p_\zeta b_{r} (\zeta_B^{+}) = D_B$ and $\p_\zeta b_{\phi} (\zeta_B^{+}) = 0$, we obtain:
\begin{equation}\label{AN.Ohm.uphi_jump}
    u_\phi(\zeta_B^+) - u_\phi(\zeta_B^-)
    = \eta_O \p_\zeta b_\phi(\zeta_B^-).
\end{equation}

The other two boundary conditions come from assuming that the magnetic field does not vary significantly at the transition between the two regions:
\begin{equation}
    b_r(\zeta_B^-) 
    = b_r(\zeta_B^+)
    = b_{rs} + D_B \zeta_B,
\end{equation}
\begin{equation}
    b_\phi (\zeta_B^-) 
    = b_\phi (\zeta_B^+)
    = b_{\phi s}.
\end{equation}

The radial magnetic field profile in the passive-field region is then given by:
\begin{equation}\label{matched.hydro.br}
    b_{r}(\zeta) = b_{r1}\zeta
    = \left[ \f{b_{rs} }{\zeta_B} + D_B 
     \right] \zeta.
\end{equation}

While the azimuthal magnetic field profile in the passive-field region is:
\begin{equation}
    b_\phi 
    = \f{1}{6\eta_O}\left( - 2 u_{\phi2} + \f{3}{2} b_{r1} \right) \zeta^3 + b_{\phi1}\zeta.
\end{equation}
where
\begin{equation}
    b_{r1} = \f{b_{rs} }{\zeta_B} + D_B,
\end{equation}
\begin{equation}
    b_{\phi 1}
    = \f{b_{\phi s}}{\zeta_B} 
    - \f{1}{6\eta_O}\left( - 2 u_{\phi2} + \f{3}{2} b_{r1} \right) \zeta_B^2.
\end{equation}

The azimuthal velocity field in the passive-field region is given by \eqref{AN.Ohm.hydro.uphi}, while in the force-free field region we use \eqref{AN.Ohm.uphi_jump} to find:
\begin{equation}
    u_\phi 
    = \f{3}{2}\left(
    b_{rs} \lvert \zeta \rvert 
    + \f{D_B}{2} \zeta^2 \right)
    + u'_{\phi 0},
\end{equation}
\begin{equation}
    u'_{\phi 0}
    = \eta_O \p_\zeta b_\phi(\zeta_B^-)
    + u_\phi(\zeta_B^-)
    - \f{3}{2}\left(b_{rs}\lvert\zeta_B\rvert
    + \f{D_B \zeta_B^2}{2} \right).
\end{equation}

To calculate the flux transport, we evaluate the gradients of the horizontal magnetic fields in the passive-field region:
\begin{equation}
    \p_\zeta b_r^-
    = \f{b_{rs} }{\zeta_B} + D_B,
\end{equation}
\begin{equation}
    \p_\zeta b_\phi^-
    = \f{1}{3\eta_O}\left( - 2 u_{\phi2} + \f{3}{2} b_{r1} \right) \zeta^2 + b_{\phi1}.
\end{equation}

This gives us the flux transport velocity as:
\begin{equation}
    u_\Psi = \eta_O \f{b_{rs} }{\zeta_B} .
\end{equation}
This shows us that the Ohm term can drive flux transport given the presence of an inclination in the poloidal field to the vertical. A large scale radial magnetic gradient also contributes to flux transport.

\subsection{Approximate analytic model when $\eta_H\gg\eta_O$}
\label{app:GO.Analytic.HallDom}

We follow the same procedure as Appendix \ref{app:GO.Analytic.OhmOnly} but with $\eta_H$ included.

\subsubsection{Purely hydrodynamic, $\beta\to\infty$ [Passive field]}
The equations become:
\begin{equation}\label{hydro:1}
    - 2 \Tilde{u}_\phi 
    =
    \f{3}{2} + D_H - D_{\nu\Sigma}
    + \left( \f{3}{2} -  D_H \right) \zeta^2 ,
\end{equation}
\begin{equation}\label{hydro:2}
    \Tilde{u}_r 
    = 0 ,
\end{equation}
\begin{equation}\label{hydro:3}
    \p_\zeta\left( \eta_{O} \p_\zeta  b_r \right)
    + \p_\zeta\left( \eta_H \p_\zeta b_\phi \right)
    = D_B \p_\zeta \left( {\eta}_{O} \right) ,
\end{equation}
\begin{equation}\label{hydro:4}
    \p_\zeta\left( \Tilde{\eta}_H \p_\zeta  b_r \right)
    - \p_\zeta \left( \Tilde{\eta}_{O} \p_\zeta b_\phi \right)
    - \p_\zeta u_\phi + \f{3}{2} b_r 
    = D_B \p_\zeta \Tilde{\eta}_{H} .
\end{equation}

From \eqref{hydro:1}, we can write
\begin{equation}\label{hydro.uphi}
    u_\phi = u_{\phi0} + u_{\phi2} \zeta^2,
\end{equation}
\begin{equation}
    u_{\phi0} 
    = -\f{1}{2} \left( 
    \f{3}{2} + D_H - D_{\nu\Sigma} \right),
\end{equation}
\begin{equation}
    u_{\phi2}
    = \f{1}{2}\left( D_H - \f{3}{2} \right).
\end{equation}

Assuming constant diffusivities, we have:\\
\eqref{hydro:3}:
\begin{equation}\label{hydro:3.1}
    \eta_O\p^2_\zeta b_r + \eta_H \p_\zeta^2 b_\phi = 0,
\end{equation}
\eqref{hydro:4}:
\begin{equation}\label{hydro:4.1}
    \eta_H\p^2_\zeta b_r - \eta_O \p_\zeta^2 b_\phi 
    - 2 u_{\phi2} \zeta + \f{3}{2} b_r
    = 0.
\end{equation}

We find $B_r$ via
\eqref{hydro:4.1}$\times\eta_H$ $+$ \eqref{hydro:3.1} $\times \eta_O$, which after some rearrangement gives us:
\begin{equation}
    \p^2_\zeta b_r
    + \f{3}{2} \left(\f{\eta_H}{\eta_O^2+\eta_H^2}\right) b_r
    = 2 u_{\phi2} \left(\f{\eta_H}{\eta_O^2+\eta_H^2}\right) \zeta.
\end{equation}
This can be solved as a second order ODE using standard techniques:
\begin{equation}
    b_{r,\text{CF}} = B_1 \cos{(\kappa \zeta)} + B_2 \sin{(\kappa \zeta)},
\end{equation}
\begin{equation}
    \kappa = \sqrt{\f{3}{2} \left(\f{\eta_H}{\eta_O^2+\eta_H^2}\right)},
\end{equation}
\begin{equation}
    b_{r,\text{PI}} = \f{4}{3} u_{\phi2} \zeta,
\end{equation}
\begin{equation}
    b_{r,\text{GS}} = b_{r,\text{CF}} + b_{r,\text{PI}}.
\end{equation}
Applying midplane boundary conditions, we obtain $B_1=0$. Hence,
\begin{equation}\label{hydro.br}
    b_{r}(\zeta) = B_2 \sin{(\kappa \zeta)} +  \f{4}{3} u_{\phi2} \zeta,
\end{equation}
\begin{equation}
    u_{\phi2}
    = \f{1}{2}\left( D_H - \f{3}{2} \right).
\end{equation}
Note that when $\eta_H=0$, $\p_\zeta^2 b_r = 0$, and
\begin{equation}
    b_r = b_{r1} \zeta,
\end{equation}
where $b_{r1}$ is a constant to be determined.\\

We find $B_\phi$ via
\eqref{hydro:3.1}$\times\eta_H$ $-$ \eqref{hydro:4.1}$\times\eta_O$, which after rearranging gives us:
\begin{equation}
    \p^2_\zeta b_\phi 
    = \f{3}{2} B_2 \left(\f{\eta_O}{\eta_O^2 + \eta_H^2}\right) \sin{(\kappa \zeta)}.
\end{equation}

When $\eta_O\neq0$,\ $\eta_H=0$:
\begin{equation}
    \p^2_\zeta b_\phi = 0,
\end{equation}
\begin{equation}
    b_{\phi}(\zeta) = b_{\phi1}\zeta.
\end{equation}
When $\eta_O,\eta_H\neq0$:
\begin{equation}\label{hydro.bphi}
    b_\phi = - B_2 \f{\eta_O}{\eta_H} \sin{(\kappa\zeta)} + b_{\phi1}\zeta.
\end{equation}
As $\eta_H\to0$, $\kappa\to0$ but the limit is such that $b_\phi\to\pm \infty$.

The flux transport in this region is given by Equation (82):
\begin{equation}\label{hydro.uPsi}
    u_\Psi = \eta_O (\p_\zeta b_r - D_B) + \eta_H \p_\zeta b_\phi.
\end{equation}

\subsubsection{Force-free magnetic field}

By the same arguments outlined in Appendix \ref{app:GO.Analytic.OhmOnly.forcefree}, we have:
\begin{equation}\label{forcefree.2}
    \p_\zeta b_\phi = 0,
\end{equation}
\begin{equation}\label{forcefree.1}
    \p_\zeta b_r - D_B = 0.
\end{equation}
With our boundary conditions at infinity, we find that throughout the force-free region:
\begin{equation}\label{forcefree.bphi}
    b_\phi = \pm b_{\phi s},
\end{equation}
\begin{equation}\label{forcefree.br}
    b_r = \pm b_{rs} + D_B \zeta,
\end{equation}
where $\pm$ stands for the sign of $\zeta$.

The absence of a current means that the magnetic field cannot diffuse.
Velocity is dtermined by the fact that the fluid is frozen into the magnetic field lines.

For the azimuthal velocity, substituting \eqref{forcefree.br} and \eqref{forcefree.bphi} to Equation (64) and integrating over $\zeta$:
\begin{equation}
    u_\phi 
    = \f{3}{2}\left(
    b_{rs} \lvert \zeta \rvert 
    + \f{D_B}{2} \zeta^2 \right)
    - D_B \eta_H 
    + u'_{\phi 0}.
\end{equation}
The radial velocity is equal to the flux transport velocity and is a constant:
\begin{equation}
    u_r = u_\Psi.
\end{equation}

\subsubsection{Two-zone model}
The height at which the magnetic pressure is equal to the thermal pressure is given by:
\begin{equation}
    \zeta_B = \sqrt{\ln{\left( \f{2}{\pi} \beta_0^2 \right)}}.
\end{equation}
We make the same assumptions and follow through the argument of Appendix \ref{app:GO.Analytic.OhmOnly.twozone} to determine the boundary conditions at $\zeta=\zeta_B$ to connect the two zones.

Two boundary conditions can be obtained from the analysis of the induction equation. Using \eqref{hydro.uPsi}, \eqref{hydro.br} and \eqref{hydro.bphi} at $\zeta=0$, we obtain:
\begin{equation}
    u_\Psi = \eta_O \left[\f{4}{3} u_{\phi2} - D_B\right] + \eta_H b_{\phi1}.
\end{equation}

The azimuthal component of the induction is more complicated because $b_r$ acts as a source term. We integrate between two heights $\zeta_1$ and $\zeta_2$ to get:
\begin{equation}
    \left[ \eta_O \p_\zeta b_\phi 
    + \eta_H ( D_B - \p_\zeta b_r )
    + u_\phi 
    \right]^{\zeta_2}_{\zeta_1}
    = 
    \f{3}{2} \int^{\zeta_2}_{\zeta_1} b_r \rmd \zeta.
\end{equation}
We use this relation between $\zeta_B^{-}$ and $\zeta_B^{+}$ to connect the two regions and neglect the width of the intermediate region, so the RHS of the equation vanishes. The force-free condition gives us $\p_\zeta b_{r} (\zeta_B^{+}) = D_B$ and $\p_\zeta b_{\phi} (\zeta_B^{+}) = 0$. Hence,
\begin{equation}
    u_\phi(\zeta_B^+) - u_\phi(\zeta_B^-)
    = \eta_O \p_\zeta b_\phi(\zeta_B^-)
    - \eta_H \p_\zeta b_r (\zeta_B^-).
\end{equation}

The two other conditions come from the two components of the equation of motion. 
Assuming that the magnetic field does not vary significantly at the transition between the two regions:
\begin{equation}
    b_r(\zeta_B^-) 
    = b_r(\zeta_B^+)
    = b_{rs} + D_B \zeta_B,
\end{equation}
\begin{equation}
    b_\phi (\zeta_B^-) 
    = b_\phi (\zeta_B^+)
    = b_{\phi s}.
\end{equation}

The radial magnetic field profile in the passive-field region is given by:
\begin{equation}\label{matched.hydro.br}
    b_{r}(\zeta) = B_2 \sin{(\kappa \zeta)} + \f{4}{3} u_{\phi2} \zeta,
\end{equation}
where
\begin{equation}
    B_2 = \f{b_{rs} + \left(D_B 
    - \f{4}{3} u_{\phi2} \right)\zeta_B }{\sin{(\kappa \zeta_B)}},
\end{equation}

The azimuthal magnetic field profile in the passive-field region is:
\begin{equation}
    b_\phi = - B_2 \f{\eta_O}{\eta_H} \sin{(\kappa\zeta)} + b_{\phi1}\zeta.
\end{equation}

The azimuthal velocity field in the passive-field region is given by \eqref{hydro.uphi}, while in the force-free field region it is:
\begin{equation}
    u_\phi 
    = \f{3}{2}\left(
    b_{rs} \lvert \zeta \rvert 
    + \f{D_B}{2} \zeta^2 \right)
    - D_B \eta_H 
    + u'_{\phi 0},
\end{equation}
\begin{equation}
\begin{aligned}
    u'_{\phi 0}
    = ~& \eta_O \p_\zeta b_\phi(\zeta_B^-)
    - \eta_H\p_\zeta b_r(\zeta_B^-)
    + u_\phi(\zeta_B^-) \\
    & 
    - \f{3}{2}\left(b_{rs}\lvert\zeta_B\rvert
    + \f{D_B \zeta_B^2}{2} \right)
    + D_B \eta_H.
\end{aligned}
\end{equation}

The gradients of the horizontal magnetic fields in the passive-field region are:
\begin{equation}
    \p_\zeta b_r^-
    = B_2 \kappa \cos{(\kappa\zeta)}
    + \f{4}{3} u_{\phi2},
\end{equation}
\begin{equation}
    \p_\zeta b_\phi^-
    = - B_2\kappa\f{\eta_O}{\eta_H}\cos{(\kappa\zeta)}
    + b_{\phi1}.
\end{equation}

The flux transport is then explicitly calculated as:
\begin{equation}
    u_\Psi = 
    \eta_H \f{b_{\phi s} }{\zeta_B}
    + \eta_O \f{b_{rs}}{\zeta_B}.
\end{equation}

\subsection{Ohmic diffusion only incorporating effects from the intermediate transition region}
\label{app:GO.Analytic.OhmOnly.Inter}

Here, we assume intermediate region to the thin, and concentrate only on the $D_B$, $b_{rs}$ and $b_{\phi s}$ source terms (neglecting Keplerian and Hydrodynamic source terms).

The density profile around $\zeta_B$ is approximated by:
\begin{equation}
    \Tilde{\rho} \simeq \f{1}{2\beta_0}\exp{(-\zeta_B x)},
\end{equation}
where $x\equiv\zeta-\zeta_B$. 
The intermediate region where $\beta\sim1$ has a typical thickness of $x\sim 1/\zeta_B$,
and our assumption of a thin transition is valid if $\zeta_B^2 \gg 1$.

The differential system is then rewritten as:
\begin{equation}\label{inter:1}
    - 2 \Tilde{u}_\phi - 2 \mathrm{e}^{\zeta_B x} \p_\zeta b_r
    =
    - 2 \mathrm{e}^{\zeta_B x} D_B ,
\end{equation}
\begin{equation}\label{inter:2}
    \f{1}{2} \Tilde{u}_r - 2 \mathrm{e}^{\zeta_B x} \p_\zeta b_\phi
    = 0 ,
\end{equation}
\begin{equation}\label{inter:3}
    \p_\zeta \left[ \Tilde{\eta}_{O} 
    \left( \p_\zeta  b_r - D_B \right)
    + u_r \right]
    = 0 ,
\end{equation}
\begin{equation}\label{inter:4}
   \p_\zeta \left (\Tilde{\eta}_{O} \p_\zeta b_\phi 
    + u_\phi \right)
    = \f{3}{2} b_r .
\end{equation}
We integrate between two heights $\zeta_1$ and $\zeta_2$ to get:
\begin{equation}
    \left[ \eta_O \p_\zeta b_\phi 
    + u_\phi 
    \right]^{\zeta_2}_{\zeta_1}
    = 
    \f{3}{2} \int^{\zeta_2}_{\zeta_1} b_r \rmd \zeta.
\end{equation}
As before, we use this relation between $\zeta_B^{-}$ and $\zeta_B^{+}$ to connect the two regions, neglecting the width of the intermediate region, so the RHS of the equation vanishes. The force-free condition is also employed where $\p_\zeta b_{r} (\zeta_B^{+}) = D_B$ and $\p_\zeta b_{\phi} (\zeta_B^{+}) = 0$. Hence,
\begin{equation}
    u_\phi(\zeta_B^+) - u_\phi(\zeta_B^-)
    = \eta_O \p_\zeta b_\phi(\zeta_B^-)
    - \eta_H \p_\zeta b_r (\zeta_B^-).
\end{equation}
and 
\begin{equation}
\begin{aligned}
    u_\Psi & = u_r + \eta_O (\p_\zeta b_r - D_B) \\
    & = - \eta_O D_B  .
\end{aligned}
\end{equation}
Thus,
\begin{equation}
    u_r = u_\Psi - \eta_O (\p_x b_r - D_B) ,
\end{equation}
\begin{equation}
    u_\phi = \Delta u_\phi 
    - \eta_O \p_x b_\phi
    ,
\end{equation}
where $\Delta u_\phi = u_\phi(\zeta_B^+) - u_\phi(\zeta_B^-) = \eta_O \p_\zeta b_\phi(\zeta_B^-)$.

We substitute these relations into \eqref{inter:1} and \eqref{inter:2}:
\begin{equation}\label{bound.inter:1}
    \eta_O \p_x b_\phi 
    - \mathrm{e}^{\zeta_B x} (\p_x b_r - D_B )  
    =
    \Delta u_\phi ,
\end{equation}
\begin{equation}\label{bound.inter:2}
    \eta_O (\p_x b_r - D_B)
    + 4 \mathrm{e}^{\zeta_B x} \p_x b_\phi
    =
    u_\Psi .
\end{equation}

We can treat them as simultaneous equations in $\p_x b_r - D_B$ and $\p_x b_\phi$:
\begin{equation}\label{simul.bound.inter:1}
    - \mathrm{e}^{\zeta_B x} (\p_x b_r - D_B)
    + \eta_O  \p_x b_\phi 
    =
    \Delta u_\phi ,
\end{equation}
\begin{equation}\label{simul.bound.inter:2}
    \eta_O (\p_x b_r - D_B)
    + 4 \mathrm{e}^{\zeta_B x} \p_x b_\phi
    =
    u_\Psi .
\end{equation}
We find $\p_x b_r$ via \eqref{simul.bound.inter:2}$\times \eta_O - $\eqref{simul.bound.inter:1}$\times 4 \mathrm{e}^{\zeta_B x} $, which after rearranging gives:
\begin{equation}
    \p_x b_r 
    =
    D_B +
    \f{
    \eta_O u_\Psi
    - 4 \mathrm{e}^{\zeta_B x} 
    \Delta u_\phi 
      }
    { \eta_O^2
    + 4 \mathrm{e}^{2\zeta_B x}
    }.
\end{equation}

We find $\p_x b_\phi$ via
\eqref{simul.bound.inter:1}$\times \eta_O - $\eqref{simul.bound.inter:2}$\times \mathrm{e}^{\zeta_B x}$, which after rearranging gives:
\begin{equation}\label{eq:HallDom.AN.dxbphi}
    \p_x b_\phi 
    = \f{
    \eta_O \Delta u_\phi 
    + \mathrm{e}^{\zeta_B x} u_\Psi }
    {\eta_O^2
    + 4 \mathrm{e}^{2\zeta_B x} }.
\end{equation}

Integrating over $x$, we compute the jump in $b_r$ across the intermediate region if $\mathrm{e}^{\zeta_B x} \gg 1$:
\begin{equation}
\begin{aligned}
    & b_r(x)-b_r(-x) \\
    \simeq & ~
    2 D_B x \\
    & + \f{1}{\zeta_B}\f{u_\Psi}{\eta_O} 
    \cdot \Bigg \{ 2\zeta_B x - \f{1}{2}\ln{\left(\f{4}{\eta_O^2} \mathrm{e}^{2\zeta_B x} 
    \right)} 
    \Bigg \} \\
    & ~
    - \f{\pi}{\zeta_B}
    \f{ \Delta u_\phi } 
    {\eta_O} .
\end{aligned}
\end{equation}

The jump condition is thus:
\begin{equation}
    \Delta b_r = - \f{\pi}{\eta_O} \f{\Delta u_\phi}{\zeta_B}
    - \ln{\left(\f{2}{\eta_O}\right)}\f{u_\Psi}{\eta_O \zeta_B}.
\end{equation}

Doing the same for $b_\phi$, we have:
\begin{equation}
\begin{aligned}
    & b_\phi(x)-b_\phi(-x) \\
    \simeq & ~
    \f{1}{\zeta_B}\Bigg\{ \f{\Delta u_\phi }{\eta_O } \Bigg\}
    \cdot \Bigg \{ 2\zeta_B x - \f{1}{2}\ln{\left(\f{4}{\eta_O^2} \mathrm{e}^{2\zeta_B x} 
    \right)} 
    \Bigg \} \\
    & ~
    + \f{\pi }{\zeta_B}
    \f{ u_\Psi} 
    {4 \eta_O} .
\end{aligned}
\end{equation}

The jump condition is thus:
\begin{equation}
    \Delta b_\phi = \f{4\pi}{\eta_O} \f{ u_\Psi}{\zeta_B}
    - \ln{\left(\f{2}{\eta_O}\right)}\f{\Delta u_\phi}{\eta_O \zeta_B} .
\end{equation}

These new boundary conditions can be injected into the two zone model such that
\begin{equation}
    b_r(\zeta_B^-) = b_r(\zeta_B^+) - \Delta b_r,
\end{equation}
\begin{equation}
    b_\phi(\zeta_B^-) = b_\phi(\zeta_B^+) - \Delta b_\phi.
\end{equation}

Hence for $b_r$, we have
\begin{equation}
    b_{r1} \zeta_B = \f{\pi}{\eta_O} \f{\Delta u_\phi}{\zeta_B}
    + \ln{\left(\f{2}{\eta_O}\right)}\f{u_\Psi}{\eta_O \zeta_B}
    + b_{rs} + D_B \zeta_B.
\end{equation}
After some algebra, we obtain
\begin{equation}
\begin{aligned}
    b_{r1} 
    = ~& \f{1}{\zeta_B\left(
    1 - \f{\pi}{2\eta_O} \right)
    - \f{1}{\zeta_B}\ln{\left(\f{2}{\eta_O}\right) } } \\
    & \times \left\{ \f{\pi b_{\phi s}}{\zeta_B^2}
    + b_{rs}
    + D_B \left[ \zeta_B - \f{1}{\zeta_B}\ln{\left(\f{2}{\eta_O}\right)} \right]
    \right\}.
\end{aligned}
\end{equation}

For $b_\phi$, we obtain 
\begin{equation}
\begin{aligned}
    b_{\phi 1}
    = & ~ \left[ 
    \zeta_B
    - \ln{\left(\f{2}{\eta_O}\right)}
    \f{1}{\zeta_B}
    \right]^{-1} \\
    & ~ \times
    \Bigg[ 
    b_{\phi s}
    - \f{\pi (b_{r1} - D_B)}{4\zeta_B}
    - \f{1}{4\eta_O} b_{r1} \zeta_B^3 \\
    & \qquad
    + \ln{\left(\f{2}{\eta_O}\right)}\f{\zeta_B}{\eta_O}
    \left[ 
    \f{3}{4}b_{r1} )
    \right]
    \Bigg]
\end{aligned}
\end{equation}

Under these new conditions, the flux transport is:
\begin{equation}
    u_\Psi = \eta_O \left[\p_\zeta \left(b_{r1}\zeta\right) - D_B\right],
\end{equation}
\begin{equation}
    u_\Psi = \eta_O \left(b_{r1} - D_B\right) .
\end{equation}

\bsp	
\bibliography{MAIN.bbl}
\label{lastpage}
\end{document}